\documentclass[]{article}
\usepackage[utf8]{inputenc} %
\usepackage[T1]{fontenc} %

\usepackage{graphicx}
\usepackage{amsmath}
\usepackage{amssymb}
\usepackage{empheq}
\usepackage[most]{tcolorbox}
\usepackage{slashed}
\usepackage{mathrsfs}
\usepackage{amsfonts}
\usepackage{braket}
\usepackage{hyperref}
\usepackage{cancel}
\usepackage{cleveref}
\usepackage{wasysym}
\usepackage{subcaption}
\usepackage[normalem]{ulem}
\usepackage{multirow}
\newcommand{\dd}{\mathop{\mathrm{d}\!}{}}

\newcommand{\deriv}[2]{\dfrac{\dd #1}{\dd #2}}

\newcommand{\OO}{\mathcal{O}}

\newcommand{\eqdef}{\equiv}%
\newcommand{\si}[1]{\mathrm{#1}}
\DeclareRobustCommand{\rchi}{{\mathpalette\irchi\relax}}
\newcommand{\irchi}[2]{\raisebox{\depth}{$#1\chi$}} %

\newcommand{\sol}{{\rm{sol}}}
\newcommand{\BH}{{\rm{BH}}}

\newtcbox{\mymath}[1][]{%
    nobeforeafter, math upper, tcbox raise base,
    enhanced, colframe=blue!30!black,
    colback=blue!30, boxrule=1pt,
    #1}

\usepackage[backend=bibtex,style=numeric-comp,sorting=none,natbib=true, url=false, doi=false, isbn=false]{biblatex} %

\usepackage{bm}
\let\vec\bm

\def\r{\tilde{r}}
\def\x{\tilde{x}}

\makeatletter

\@addtoreset{equation}{section}

\makeatother

\begin{document}
\setcounter{page}{0}
\thispagestyle{empty}

\parskip 3pt

\font\mini=cmr10 at 2pt

\begin{titlepage}
~\vspace{1cm}
\begin{center}

	\vspace*{-.6cm}

	\begin{center}

		\vspace*{1.1cm}

		{\centering \Large\textbf{Interplay between Black Holes and Ultralight Dark Matter: Analytic Solutions}}

	\end{center}

	\vspace{0.8cm}
	{\bf Bruno Bucciotti$^{a,b}$, Enrico Trincherini$^{a,b}$}

	\vspace{1.cm}
	
	${}^a\!\!$
	{\em  Scuola Normale Superiore, Piazza dei cavalieri 7, 56126 Pisa, Italy}
		
	\vspace{.3cm}

	${}^b\!\!$
	{\em INFN, Sezione di Pisa, Largo B. Pontecorvo, 3, 56127 Pisa, Italy}
			
\end{center}

\begin{abstract}

Dark matter (DM) can consist of a scalar field so light that DM particles in the galactic halo are best described by classical waves. We investigate how these classical solutions are influenced by the presence of a non-rotating supermassive black hole at the center of the galaxy, using an analytical, albeit approximate, approach. 

Relying on this analytic control, we examine the consequences of imposing causal boundary conditions at the horizon, which are typically overlooked.
First, we examine the scenario where the backreaction of dark matter can be neglected. The scalar field decays like a power law at large distances, thus endowing the black hole with “hair”. We derive solutions for the field profile over a wide range of parameters, including cases with rotating dark matter. As a by-product, we extract the dynamical Love numbers for scalar perturbations. Next, we determine the spectrum of bound states and their behaviour. 

Finally, we incorporate the self-gravity of the scalar field, with a focus on the situation where dark matter forms a soliton (boson star) at the center of the galaxy. We derive an analytical expression for the soliton at every distance from the center. With a solution that remains applicable even at horizon scales, we can reliably compute the accretion rate of the black hole.

\end{abstract}

\end{titlepage}

\tableofcontents
\section{Introduction}

Quoting Edward Witten~\cite{Adventures}: \textquotedblleft One side of what theoretical physicists do is to try to understand the fundamental equations of nature, and the other side is to try to solve the equations in different situations and work out predictions for what will happen.\textquotedblright

While this paper discusses a new hypothetical constituent of the Universe in the form of an ultralight scalar boson, our focus will be on the latter aspect. The equation describing this boson is actually quite simple: a massive scalar field with no self-interactions, minimally coupled to gravity. However, our specific interest lies in a particular class of solutions that describe a self-gravitating, spherically-symmetric, stationary configuration of the scalar in the presence of a black hole (BH) at the center. Furthermore, we will adopt an analytical, albeit approximate, approach to study such solutions.

The main motivation for considering ultralight scalar fields arises from the existence of dark matter (DM). Within a wide mass range of $10^{-22}-10^{-21}$ eV to $10$ eV, if we assume that all of the dark matter originates from the light scalar and take into account the density of dark matter in the galactic halo, the occupation number of the particles becomes so large that they behave as oscillating classical fields. Hence, this scenario is commonly referred to as wave dark matter (WDM)~\cite{Baldeschi:1983mq,Turner:1983he,Press:1989id,Sin:1992bg,Peebles:2000yy,Hu:2000ke,Lesgourgues:2002hk, Amendola:2005ad,Suarez:2011yf,Rindler-Daller:2011afd,Schive:2014dra,Hui:2016ltb,Hui:2021tkt}. 
The most compelling example within this category is the QCD axion, followed perhaps by axion-like particles (ALP) predicted by string theory~\cite{Svrcek:2006yi,Arvanitaki:2009fg,Halverson:2017deq,Bachlechner:2018gew}.

At the lighter end of the mass spectrum, the scalar is known as fuzzy dark matter (FDM). In particular, FDM was proposed to address certain issues pertaining to the small-scale features observed in simulations conducted using conventional cold dark matter models~\cite{Hu:2000ke}.

In the presence of gravity, a free scalar field can support itself against gravitational collapse by the quantum property that particles cannot be localized beneath distances of the order of their Compton wavelength. The resulting (quasi-)stationary solutions of the field equations of motion are commonly known as boson stars (for a complex scalar field)~\cite{Kaup:1968zz,Ruffini:1969qy,Friedberg:1986tq,Lee:1991ax,Liebling:2012fv}, oscillatons (for a real scalar)~\cite{Seidel:1991zh,Copeland:1995fq,Urena-Lopez:2001zjo,Page:2003rd,Visinelli:2021uve}, or solitons (the term we will use in the following). There exists an upper limit to the mass of a soliton $M_{\rm sol}$, as a function of the scalar field mass, in order to prevent collapse into a black hole. Simulations~\cite{Schive:2014dra,Schive:2014hza} indicate that fuzzy dark matter halos typically exhibit a solitonic core surrounded by a halo with a Navarro–Frenk–White (NFW) profile~\cite{Navarro:1995iw}. Numerous papers have compared this expectation against observational data~\cite{Marsh:2015wka,bar_galactic_2018,Bar:2019bqz,bar_looking_2019,Desjacques:2019zhf,Bar:2021kti}. However, our work deviates from this approach and instead addresses a different question: how is the wave dark matter soliton in a galactic halo influenced by the presence of a supermassive black hole (SMBH) at the center of the soliton? For this reason, we will take a broader perspective than fuzzy dark matter and consider the mass $\mu$ of the scalar as a free parameter within the WDM range.

An important element, once additional matter in the form of a black hole is added to the scalar condensate, is the boundary condition at the black hole horizon. In much of the literature, which primarily focuses on distances significantly larger than the black hole horizon $r_{\rm BH}$, its effect is often neglected. While the causal solution should be outgoing (entering the horizon), in principle the time-reversed mode also exists. When solving the equations from infinity, one generically encounters a superposition of both modes at the horizon. Prohibiting the non-causal mode of the scalar field could impact the solution, potentially even at distant locations. Numerical control is challenging due to the presence of the horizon and the highly oscillating nature of the solution. Therefore, as mentioned earlier, we will instead employ an analytic approach. For other investigations on how a black hole modifies the soliton see~\cite{Urena-Lopez:2002nup,Barranco:2011eyw,Brax:2019npi,Chavanis:2019bnu,bar_looking_2019,Davies:2019wgi,cardoso_parasitic_2022}. Black hole accretion of diffuse scalar is studied instead in~\cite{Cruz-Osorio:2010nua,Lam,Clough:2019jpm}.    
\bigskip

The structure of the paper is as follows. We begin Section~\ref{sec:no_self_gravity} discussing the case where the backreaction of dark matter can be neglected, and the equation for the scalar field reduces to the Klein-Gordon equation in a Schwarzschild background. In such cases there exists more literature, dating back to Starobinskii~\cite{Starobinskii:1973hgd} and briefly summarized in~\cite{Hui:2021tkt}, and the exact solution in terms of a special function, known as the confluent Heun function~\cite{Bezerra:2013iha}, has been derived. However, manipulating this exact solution is challenging. Therefore, a systematic exploration of the resulting hairy black hole solutions was conducted in~\cite{Lam} for different values of the dark matter Compton wavelength $\hbar/ \mu$ compared to $r_{\rm BH}$, in the case where dark matter has no angular momentum. Contrary to solitonic solutions, these profiles decay like a power law at large distances, thus they endow the black hole with \textquotedblleft hair\textquotedblright\, that is then matched to the halo. Notice that no-hair theorems do not apply because the scalar field is not static. 

We will address this scenario using different techniques that provide better control over the approximate solutions and allow for the inclusion of spinning dark matter. In the large mass (or particle) limit, $\mu r_{\rm BH} \gg 1$, the method of uniform approximations~\cite{Berry}, which generalizes the standard WKB approach, will be employed. In the opposite (wave) limit  $\mu r_{\rm BH} \ll 1$, where both WKB and uniform approximations fail, the approximate solution will be derived by employing the technique of boundary layer theory. 
The particle limit solution will be sensitive to the boundary condition at the horizon at all distances. On the other hand, in the wave limit, the boundary conditions become negligible within a few black hole radii. In addition our solution enables us to compute the dynamical response coefficients (Love numbers) for a massive scalar field in the small mass limit. Where they overlap, our results are consistent with~\cite{Charalambous:2021mea,Kehagias:2022ndy}.

The intermediate mass range  $\mu r_{\rm BH} \sim 1$, with non-zero angular momentum, poses additional challenges and will be thoroughly discussed in Appendix \ref{app:intermediate_mu}. This range is particularly intriguing as it pertains to cases where the Compton wavelength of the light scalar field is comparable to the black hole horizon, resulting in a faster growth rate of the superradiance instability. To investigate this instability, our computations would need to be generalized to a Kerr background. However, even for non-spinning black holes, our analytic formulas can be compared with the results obtained in~\cite{Hui:2022sri} by numerically evaluating the exact Heun function solution.

Finally, we end Section~\ref{sec:no_self_gravity} by discussing solitons (of mass $M_{sol}$) at the center of the dark matter halo whose dynamics is dominated by the SMBH, $M_{\rm BH}/M_{\rm sol} \gg 1$. These solutions are similar to the previously discussed hairy black hole profiles up to a length scale $r\lesssim r_{\rm{BH}}/|\gamma|$, where $\mu\gamma$ is the binding energy of a single scalar particle. At greater distances, the solution is exponentially suppressed. A second difference is that the frequency has a tiny imaginary part related to the inverse time of absorption of the soliton, which is meta-stable. We will show that causal boundary conditions at the black hole horizon impose the upper bound $\mu r_{\rm{BH}}\ll 1$ for any black hole dominated soliton. When this bound is satisfied, causal boundary conditions do not need to be explicitly checked.
\smallskip

More interesting is the case where the soliton dominates the dynamics, at least far from the SMBH. This can happen in the opposite limit, $M_{\rm BH}/M_{\rm sol} \ll 1$, and it will be discussed in Section \ref{sec:DMsoliton_domination}.
The soliton mass can be related to the halo mass via the numerically extrapolated soliton-halo relation, however we will simply keep the soliton mass as a free parameter since we do not model the halo. Neglecting the halo will not significantly affect the solution, as long as the soliton dominates over the halo density.
In this scenario a natural question arises: can the soliton dominate the dynamics at all length scales, even down to the size of the black hole's event horizon? As we will see, our analysis reveals that this is never the case. Regardless of how small  $M_{\rm BH}$ is, there always exists a length scale, denoted as $r_{\rm{e}}$, where the effect of the black hole becomes parametrically equal to that of the soliton.
Notably, this length scale satisfies the condition $r_{\rm BH} \ll r_{\rm{e}} \ll r_{\rm sol}$, where $r_{\rm sol}$ corresponds to the characteristic size of the soliton.
For $r \gg r_e$ the self-gravity dominated solution was studied both numerically~\cite{Schive:2014dra,Schive:2014hza,Davies:2019wgi,cardoso_parasitic_2022} and analytically~\cite{Tod_1999}, while for $r \ll r_e$ it reduces to the one discussed by Hui et al.~\cite{Lam} for dynamics dominated by the SMBH. The two solutions will be matched at $r\sim r_e$ by imposing continuity. We will discover that soliton stability under gravitational collapse imposes the upper bound $\mu r_{\rm{BH}}\ll \frac{M_{\rm{BH}}}{M_{\rm{sol}}}(\ll 1)$ for any self-gravity dominated soliton. We will conclude that also these kinds of solitons are only possible in the (very) small mass regime $\mu r_{\rm{BH}}\ll 1$, when causal boundary conditions are unimportant.
\medskip

{\it Conventions}: We will set $\hbar = c= 1$ throughout the paper, together with $r_{\rm BH} = 2G M_{\rm BH} = 1$ within Section~\ref{sec:no_self_gravity}. However, we will restore this value in a few key expressions and in section~\ref{sec:DMsoliton_domination} for the sake of clarity.

\section{Black hole domination}
\label{sec:no_self_gravity}

Our starting point is a complex scalar field of mass $\mu$ in a black hole background. Minimal coupling leads to the equation of motion
\begin{equation}
\label{eq:differential_equation_box}
\square\phi-\mu^2\phi = 0
\end{equation}
where the background metric is strictly Schwarzschild and gravitational backreaction is assumed to be negligible, postponing the discussion of a self-gravitating scalar.
\smallskip

Exact solutions to this equation are known in terms of the confluent Heun function. However, these solutions are not transparent or easily understandable\footnote{Besides, software like Mathematica has some difficulties in handling and plotting this function.}. The confluent Heun function is a special function that can be quite complex, making it difficult to gain insight into the behavior of the solutions just by examining the function itself. To give an example, the exact solution does not provide an explicit relation between the density of dark matter at large distance and the density near the horizon.
Progress in this direction was achieved by Bonelli et al.\ in~\cite{Bonelli:2022ten}, where they express these connection formulas in terms of a series. While their expressions are in principle exact, our goal is to find instead an {\it approximate} solution for the scalar background, valid at all distances, which, on the other hand, is expressed in terms of significantly simpler formulas. One of the results of our analysis is the identification of the different length scales that affect the field profile as a function of distance. These length scales determine where the field's behaviour undergoes significant changes. However, these length scales are not readily apparent when considering only the Heun function, emphasizing the need for alternative approaches with simpler expressions. Finally, we extend the analysis of Hui et al.~\cite{Lam} by discussing also the case of spinning dark matter.

\bigskip

The first step is to specify the time and angular dependence of the field. We look for solutions with definite frequency $\omega$ and angular momentum $l,m$. Restricting to one frequency is particularly natural because the equation is linear. Barring bound states, the field must have energy at least equal to its own mass, and as a first approximation we will neglect excited states thus setting $\omega=\mu$. The implication is that the energy density of the field $\sim|\partial_t\phi|^2$ is independent of time and so is the backreaction on the metric (which for now we take to be negligible).
\smallskip

A real scalar field would obey the same differential equation but its time dependence would be $\sim\cos(\omega t)$. The energy density $\rho\simeq \frac{1}{2}\left(\dot\phi^2+\mu^2\phi^2\right)$ would not vary with time, but the other terms in the energy-momentum tensor would, leading to a small mixing with higher harmonics and a slow decay of the field due to gravitational radiation.
\medskip

The usual separation of variables ansatz
\begin{equation}
\phi \rightarrow e^{-i\omega t} Y_{l,m}(\theta,\phi) \phi(r)
\end{equation}
brings~\ref{eq:differential_equation_box} to the form (putting $r_{\rm{BH}}=1$)
\begin{equation}
\label{eq:eq_diff}
r(r-1)\deriv{}{r}\left(r(r-1)\deriv{\phi}{r}\right)+V(r)\phi=0,\quad
V(r) \eqdef (\omega^2-\mu^2)r^4+\mu^2r^3-l(l+1)r(r-1)
\end{equation}
The form of the derivative operator that acts on $\phi$ suggests the change of variable
\begin{equation}
\label{eq:def_rtilde}
\r \eqdef \ln\left(1-\frac{1}{r}\right),\qquad
\dd\r = \frac{\dd r}{r(r-1)}
\end{equation}
because the derivative term becomes $\deriv{^2\phi}{\r^2}$. The equation takes the form of a 1d Schrödinger equation where $\r$ goes from $-\infty$ being the horizon to $\r=0^-$ being large physical distances. $\r$ will only play a formal role in what follows: when talking about distances, we will always mean $r$. This definition will prove convenient for our purposes because the derivative term of~\ref{eq:eq_diff} will not change once angular momentum is included. We stress that $\r$ is {\it not} the tortoise coordinate usually employed to analyze this kind of problems. From our perspective, this change of variables is more natural because it is the standard trick that one uses to eliminate the first order derivative in a second order differential equation. This step will allow us to employ WKB methods to look for a solution of~\ref{eq:eq_diff}.
\bigskip

Most of the literature assumes that, if we are only interested in $r\gg r_{\rm{BH}}$, we can replace $(r-1)$ with $r$ in equation~\ref{eq:eq_diff} and the horizon can be entirely neglected. However, as pointed out by Hui et al.~\cite{Lam} and Baumann et al.~\cite{Baumann:2019eav}, the role of the horizon scale is two-fold. Besides trivially entering equation~\ref{eq:eq_diff}, $r_{\rm{BH}}$ is also a boundary where we have to impose boundary conditions on $\phi$.
\smallskip

To be more explicit, we will see that the two linearly independent solutions of~\ref{eq:eq_diff} are waves decaying at infinity, one infalling and one outgoing. One could imagine setting up a scattering experiment by choosing the amplitude of the infalling wave at infinity, the reflected outgoing wave being completely fixed. What picks the amplitude of the outgoing wave? The answer is the causal boundary condition we impose at the horizon.
\smallskip

Without this boundary condition, the solution will generically be inaccurate at all distances (if one includes other interaction, such as self-gravitation of dark matter, the solution will be affected only if the black hole is dominating the dynamics). This absence of decoupling between black hole horizons and large scales is not unheard of: it is reminiscent of the black hole Love numbers which, although defined in terms of the behaviour of the field at large distances, are tuned to zero by the presence of the horizon. We will actually compute such tidal response coefficients in equation~\ref{eq:love_numbers}.
Our goal in the second part of the paper will be to check the effects of this boundary condition and develop approximations for the field profile at different distances.

\subsection{Large $\mu r_{\rm{BH}}$ regime}
\label{subsec:large_mass}
When attempting to find a solution for a challenging second-order differential equation, a very natural approach is to try the WKB approximation. Despite its high accuracy, the WKB approximation is known to be ineffective near turning points, which are locations where the particle would classically come to a halt. To address this, we begin by determining the range of the parameter space $(\mu,l)$ where the WKB approximation remains valid throughout. This will translate into a lower bound on $\mu$.
\bigskip

Looking at equation~\ref{eq:eq_diff}, the turning points are given by $V(r)=0$. We then begin our discussion by studying the sign of $V$ as a function of $r$. First,
\begin{equation}
V(1) = +\mu^2,\qquad V(r\rightarrow\infty) = +\infty
\end{equation}
are both positive. However, $V(r)$ can develop two zeros at $r>1$ and turn negative between them. Solving for $V(r)=0$ we get
\begin{equation}
\label{eq:def_r12}
r_{1,2} \eqdef \frac{l^2+l\mp\sqrt{(l+l^2)(l+l^2-4 \mu ^2)}}{2 \mu ^2}
\end{equation}
Thus there are turning points for small $\mu$ and non-zero $l$. To be more precise, only the mass range
\begin{equation}
\label{eq:critical_mass}
\mu\ge\mu_c \eqdef \frac{\sqrt{l+l^2}}{2}
\end{equation}
is free of turning points. The location of these points is well approximated by
\begin{equation}
\label{eq:def_r12_approx}
r_1 \simeq 1+\frac{\mu^2}{l+l^2},\qquad 
r_2 \simeq \frac{l+l^2}{\mu^2}
\end{equation}
in the small mass regime. We now assume $\mu\gg\mu_c$ and apply the WKB approximation, leaving the small $\mu r_{\rm{BH}}$ case for later.

\subsubsection{Uniform approximations}
Assuming $\mu\gg\mu_c(l)$ we could write down the WKB solution of equation~\ref{eq:eq_diff} in the absence of turning points, but this would turn out to be inaccurate for $l=0$. To have proper control of the approximated solution, we will rather derive the WKB approximation and the condition for its validity. Another motivation for taking this route is because it allows us to introduce the framework of uniform approximations (which vastly generalizes WKB) in a simple setting. We will use this more general framework in appendix~\ref{app:intermediate_mu}, where we will deal with two turning points of the potential close to each other. A comprehensive reference for uniform approximation techniques is \cite{Berry}.
\bigskip

The key insight of uniform approximations is to substitute~\ref{eq:eq_diff} with a \textquoteleft similar\textquoteright\, differential equation for a new function $f(x)$, defined over a new domain $x\in(1,\infty)$. We want the new differential equation to be explicitly solvable and to have the same singularity structure (i.e. zeros and poles inside the domain of interest) as the original one~\ref{eq:eq_diff}. We will call this the \textquoteleft similarity conditions\textquoteright. Under these conditions, we will be able to write down an approximate solution of~\ref{eq:eq_diff} in terms of $f(x)$ by suitably \textquotedblleft stretching\textquotedblright\, $x$ into $r$.
\smallskip

We choose $f(x)$ to obey
\begin{equation}
\label{eq:eq_diff_wkb}
(x-1)\deriv{}{x}\left((x-1)\deriv{f}{x}\right)+\Gamma(x)f(x)=0
\end{equation}
with $\Gamma(x)\equiv1$.\footnote{As explained in~\cite{Berry}, a constant $\Gamma=\pm1$ leads to WKB, while a linear $\Gamma$ describes a potential with a single turning point. In this case, this technique gives an approximate solution in terms of Airy functions which is valid both close and far from the turning point, sidestepping the usual WKB connection formulas. Appendix~\ref{app:intermediate_mu} will employ a generalization of these techniques.} To simplify the derivative term, we define (similarly to~\ref{eq:def_rtilde})
\begin{equation}
\label{eq:def_xtilde}
\x \eqdef \ln(x-1),\qquad
\dd\x = \frac{\dd x}{x-1}
\end{equation}
Since the derivative term has the same zeros in both~\ref{eq:eq_diff} and~\ref{eq:eq_diff_wkb}, and both $V(r)$ and $\Gamma$ are always positive, we can expect the solutions of~\ref{eq:eq_diff_wkb} to be a slight deformation of those of~\ref{eq:eq_diff}.
\smallskip

We formalize this idea by introducing an unknown function $\x(\r)$, which will quantify this stretching
\begin{equation}
\label{eq:uniform_solution_ansatz}
\phi(\r) \eqdef \left(\deriv{\x}{\r}\right)^{-\frac{1}{2}} f(\x(\r))
\end{equation}
Upon substitution in \ref{eq:eq_diff} and making use of \ref{eq:eq_diff_wkb}, we obtain an equation for $\x(\r)$
\begin{equation}
\label{eq:def_uniform_V}
V = \left(\deriv{\x}{\r}\right)^2 \Gamma - \left(\deriv{\x}{\r}\right)^{1/2}\frac{\dd^2}{\dd\r^2}\left(\deriv{\x}{\r}\right)^{-1/2}
\end{equation}
where $f$ has dropped out and the second term is a schwarzian derivative we wish to neglect. This term blows up whenever the singularities of the two differential equations don't match, justifying the similarity condition requirement.
\medskip

We are thus led to a simple differential equation for $\x(\r)$ together with a condition for the validity of the approximation
\begin{equation}
\label{eq:uniform_smallness_condition}
\deriv{\x}{\r} = \left(\frac{V(r)}{\Gamma(x)}\right)^{\frac{1}{2}},\qquad
\epsilon\eqdef\bigg\lvert\frac{1}{V}\left(\deriv{\x}{\r}\right)^{1/2}\frac{\dd^2}{\dd\r^2}\left(\deriv{\x}{\r}\right)^{-1/2}\bigg\rvert\ll1
\end{equation}
For $l=0$ we can easily solve this, getting $\x(\r(r)) = 2 \mu \left(\sqrt{r}-\coth ^{-1}\left(\sqrt{r}\right)\right)$ which asymptotes to $2\mu\sqrt{r}+\OO(r^{-1/2})$ for large $r$ and $\mu\left(\ln(r-1)+2-2\ln2\right)+\OO(r-1)$ when $r\simeq1$.

Given the solutions $e^{\pm i\x}$ of~\ref{eq:eq_diff_wkb} and~\ref{eq:uniform_solution_ansatz}, we obtain the sought field profile
\begin{equation}
\label{eq:large_mass_matching}
\phi_{l=0}(r)\simeq e^{\pm 2i\mu(1-\ln2)}(r-1)^{\pm i\mu}\leftrightarrow \frac{1}{r^{3/4}}e^{\pm 2i\mu\sqrt{r}}
\end{equation}
where the $\leftrightarrow$ symbol indicates that we matched the solution close to the horizon with the one at large distance. We emphasize that this matching is straightforward because the approximation is valid in the whole domain.

Subleading terms in the $r\rightarrow\infty$ expansion become comparable with the leading contribution at $r=1$. Causality at the horizon then selects the purely infalling wave $(r-1)^{-i\mu}$, thus enforcing the behaviour $\frac{e^{-2i\mu\sqrt{r}}}{r^{3/4}}$ at large distances\footnote{Had we chosen to work with a real scalar, the solutions close to the horizon would be $\phi_\pm\propto\cos(\mu\ln(r-1)\pm\mu t)$. Then $\phi_+$ would have been the causal wave.}. This solves the connection problem completely.
\bigskip

Let's draw some conclusions. We observe that, for all $l$ (up to subleading WKB orders), the amplitude of the profile is completely controlled by the overall factor $(d{\x}/d{\r})^{-\frac{1}{2}} = V^{-\frac{1}{4}}$ which has a very simple analytic expression. All complications reside in the less important phase of the wave.%

We now compute $\tilde x$ for $l>0$. The integration of~\ref{eq:uniform_smallness_condition} gives
\begin{equation}
    \tilde x = \int^r \frac{\sqrt{\mu^2r^3-l(l+1)r(r-1)}}{r(r-1)} \dd r
\end{equation}
which is hard to evaluate analytically. While a numerical solution is straightforward, let us observe that when $l=0$ we are back to the previously studied case, up to a correction which is small uniformly when $r\ge1$. Indeed $\max\frac{l(l+1)r(r-1)}{\mu^2r^3}=\frac{l(l+1)}{4\mu^2}$ which is $\le1$ when $\mu\ge \frac{\sqrt{l+l^2}}{2}$, and quickly becomes $\ll1$ in the large $\mu$ regime. Thus, upon expansion,
\begin{align}
    \tilde x\simeq 2 \mu \left(\sqrt{r}-\coth ^{-1}\left(\sqrt{r}\right)\right) - \frac{1}{2\mu} \int^r \frac{l (l+1)}{r^{3/2}} =\\
    = 2 \mu \left(\sqrt{r}-\coth ^{-1}\left(\sqrt{r}\right)\right) + \frac{l(l+1)}{\mu\sqrt{r}}
\end{align}
Higher orders are as just easy to compute. They do not affect the asymptotic expansion of $\phi$ at $r\simeq1,\infty$ but they increase the accuracy of the solution at finite $r$.
Notice that all the arbitrary constants we are dropping upon integration correspond to trivial constant phase shifts. Plugging into equation~\ref{eq:uniform_solution_ansatz} we get
\begin{empheq}[box={\mymath[colback=gray!10, sharp corners]}]{equation}
    \label{box:large_mass_full}
    \phi_l(r)\simeq \frac{1}{(\mu^2r^3-l(l+1)r(r-1))^{1/4}}e^{-2i\mu\left(\sqrt{r}-\coth ^{-1}\left(\sqrt{r}\right)\right)+\frac{l(l+1)}{\mu\sqrt{r}}}
\end{empheq}
with (near-horizon $|$ large-distances) behavior independent of $l$
\begin{empheq}[box={\mymath[colback=gray!10, sharp corners]}]{equation}
    \label{box:large_mass_matching}
    (r-1)^{\pm i\mu}\quad \bigg| \quad \frac{1}{r^{3/4}}e^{\pm 2i\mu\sqrt{r}}
\end{empheq}

We are also in the position to check $\epsilon\ll1$ given the explicit $\x$ we got. A direct computation reveals the condition
\begin{equation}
\epsilon = |-\frac{1}{\mu^2r^3}\frac{3}{16}(r-1) (r+3)|\ll1
\end{equation}
which, as anticipated, puts a lower bound $\mu\gg\frac{2}{3\sqrt{3}}\simeq0.4$ on the $l=0$ case. This reproduces the result obtained by Hui with a different technique.
\bigskip

\begin{figure}
	\centering
	\begin{subfigure}{0.49\textwidth}
		\centering
		\includegraphics[width=1.2\linewidth]{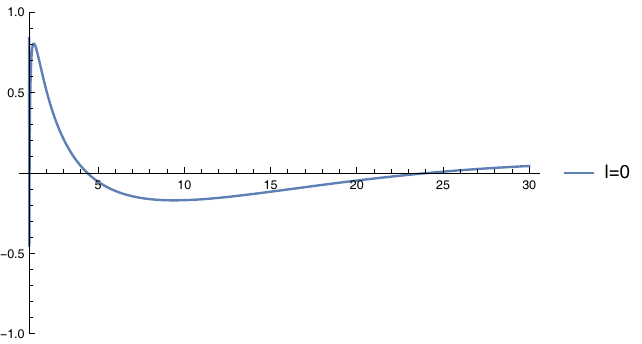}
		\caption{$\mu=0.5$. Real part}
	\end{subfigure}
	\begin{subfigure}{0.49\textwidth}
		\centering
		\includegraphics[width=1.2\linewidth]{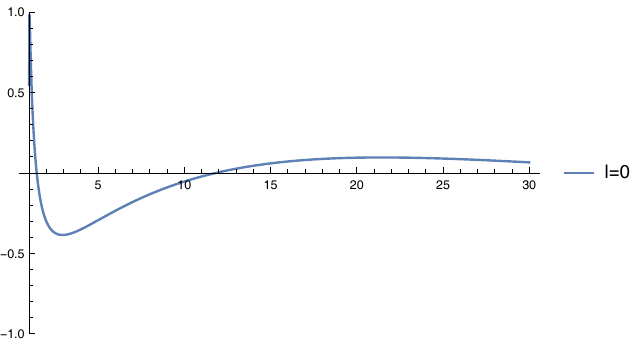}
		\caption{$\mu=0.5$. Imaginary part}
	\end{subfigure}
	\centering
	\begin{subfigure}{0.49\textwidth}
		\centering
		\includegraphics[width=1.2\linewidth]{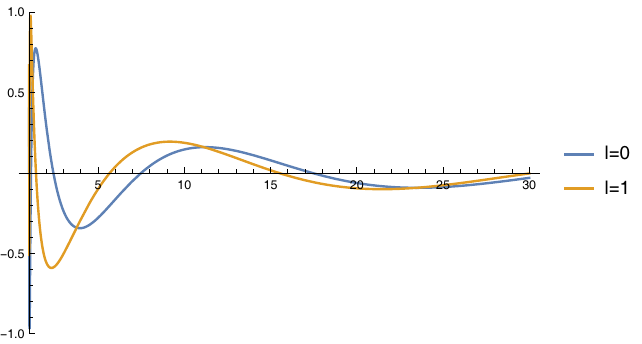}
		\caption{$\mu=1.0$. Real part}
	\end{subfigure}
	\begin{subfigure}{0.49\textwidth}
		\centering
		\includegraphics[width=1.2\linewidth]{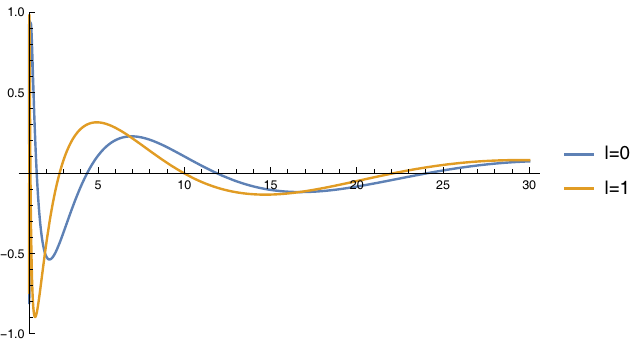}
		\caption{$\mu=1.0$. Imaginary part}
	\end{subfigure}
	\centering
	\begin{subfigure}{0.49\textwidth}
	\centering
	\includegraphics[width=1.2\linewidth]{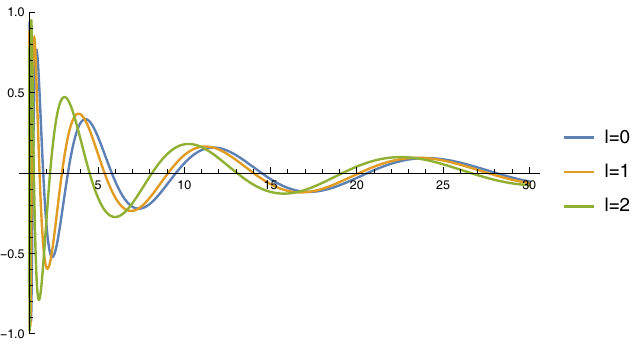}
	\caption{$\mu=2.0$. Real part}
	\end{subfigure}
	\begin{subfigure}{0.49\textwidth}
	\centering
	\includegraphics[width=1.2\linewidth]{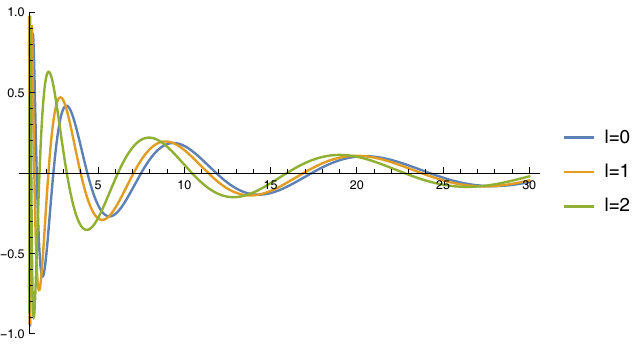}
	\caption{$\mu=2.0$. Imaginary part}
	\end{subfigure}
	\caption{Plot of our large mass approximation of $\phi$~\ref{box:large_mass_full} for different values of $\mu r_{\rm{BH}}$ and $l$. The amplitude is normalized to $1$ at the horizon.}
	\label{fig:mularge}
\end{figure}

How do we interpret this bound? In some sense, it comes from the significant smallness of the potential term $\mu^2r^3$ close to the horizon at small $\mu$: the horizon is behaving similarly to a turning point\footnote{A scattering problem with energy barely greater than the top of the potential manifests a similar problem, which can be phrased in terms of complex turning points close to the real axis.}. In contrast, when $l\ge1$, the potential has developed zeros for $\mu$ even larger than $0.4$, since $\mu_c|_{l=1}\simeq0.7$: the breakdown takes place already at the $\mu_c$ we computed in the previous subsection.
We plot our approximate solution in figure~\ref{fig:mularge}. The accretion rate can be computed to be in agreement with ref.~\cite{Lam}

The lower bound $\mu\gg0.4$ compels us to study the complementary small $\mu$ regime $\mu\ll1$, which we will tackle in the next subsection.

\begin{figure}
	\centering
	\includegraphics[width=0.6\linewidth]{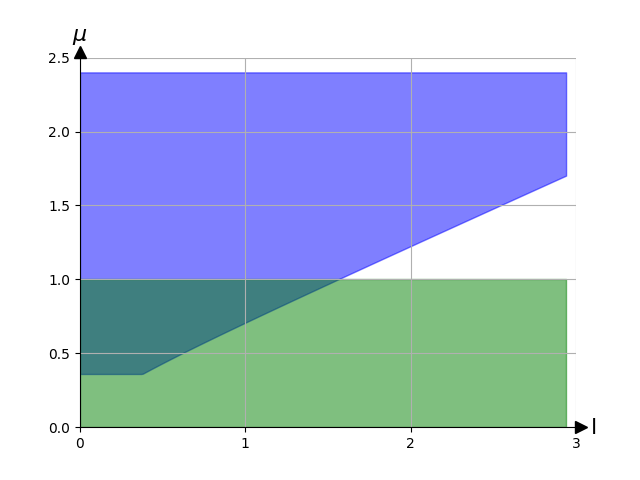}
	\caption{The large mass approximation is valid in the blue area, while we can expand for $\mu\ll1$ in the green area. Already at $l\ge2$ there exists a region beyond the scope of both approximations, which we cover in the appendix. This region enlarges if we don't want to approach the boundary of the blue and green areas, where our approximations become unreliable.}
	\label{fig:validityofapprox}
\end{figure}
\bigskip

We anticipate that the connection problem will be solved by
\begin{equation}
r=r_{\rm{BH}}:\;\phi_{in},\qquad
r\rightarrow\infty:\;\sim \mu^{-3/2-2l}\phi_{in} + \mu^{-3/2-2l}\phi_{out}
\end{equation}
up to $\OO(1)$ coefficients.
\medskip

However, as made clear by figure~\ref{fig:validityofapprox}, for $l\ge2$ there is a mass range not covered by either approximation. This will be dealt with in the appendix.

We anticipate that for $l$ small, say $l=2,3$, the connection problem in this regime is approximately solved by
\begin{equation}
r=r_{\rm{BH}}:\;\phi_{in}\qquad
r\rightarrow\infty:\;\sqrt{2}\phi_{in} + \phi_{out}
\end{equation}

\subsection{Small $\mu r_{\rm{BH}}$ regime}
\label{subsec:small_mass}
Given the failure of WKB (and uniform approximations) in the small $\mu r_{\rm{BH}}$ parameter range, we now look for a different kind of approximate solution to~\ref{eq:eq_diff} for the case $\mu r_{\rm{BH}}\ll 1$. The approach will be to neglect all terms proportional to $\mu$ in the differential equation and then identify   the $r$ domain where this is valid. We will then develop separate approximations where this assumption will fail. The upshot will be a subdivision of the domain in a near-horizon region, an intermediate region and a far region.

We refer to~\cite{Lam} for the $l=0$ case, which agrees with our result. Our main contributions are the extension of the computation to all $l$ and a more careful treatment of the matching of $\phi$ at different distances, which Hui et al.\ carried out simply by invoking continuity and derivability of the solution \textquoteleft close\textquoteright\, to the breakdown point. The technique we employ goes under the name of boundary layer theory (see the book by White \cite{White}), but we will not assume any prior knowledge of the topic.
\bigskip

Since we work in the $\mu\ll1$ approximation, we can expect $\mu^2r^3$ to be negligible in equation~\ref{eq:eq_diff}. This is true unless the $r(r-1)$ term coming from the angular momentum barrier becomes comparable or smaller. This happens in two regions, where we will instead approximate the $r$ dependence:
\begin{equation}
\label{eq:boundary_layers}
\mu^2r^3\sim r(r-1)\rightarrow 1<r<1+\mu^2,\quad r>1/\mu^2
\end{equation}
The attentive reader will recognize $1+\mu^2$ and $1/\mu^2$ as the zeros of $V(r)$ when $\mu\ll1$. The issue we had in WKB with the zeros of $V(r)$ has an avatar here as well.
\medskip

Let us begin with the intermediate $r$ region $1+\mu^2<r<1/\mu^2$, where $\mu^2 r^3$ is negligible. The solution can be written explicitly as
\begin{equation}
\label{intermed}
\phi_{\rm intermed}\simeq c_1 r^l \, _2F_1(-l,-l;-2 l;1/r)+\frac{c_2}{r^{l+1}} \, _2F_1(l+1,l+1;2 l+2;1/r)+\OO(\mu^2)
\end{equation}
where $_2F_1$ is a hypergeometric function.
\medskip

We now come to what we will call the two {\it boundary layers}: the near horizon region and the far region, defined by~\ref{eq:boundary_layers}. In the near horizon region we can solve the approximate equation
\begin{equation}
(r-1)\deriv{}{r}\left((r-1)\deriv{\phi}{r}\right)+\mu^2\phi=0
\end{equation}
which results in
\begin{equation}
\phi_{\rm n.h.} \simeq (r-1)^{-i\mu}(1+\OO(r-1))
\end{equation}
where we already imposed causality. Notice that we neglected the angular momentum part of the differential equation, but this is consistent if we don't keep track of $(r-1)^{-i\mu+1}$ terms in the solution. Boundary layer theory also gives a rigorous criterion to systematically decide which terms one should keep, but it is just as easy to verify this a posteriori.
\smallskip

Finally, the far region is controlled by
\begin{equation}
r^2\deriv{}{r}\left(r^2\deriv{\phi}{r}\right)+\left(\mu^2r^3-l(l+1)r^2\right)\phi=0
\end{equation}
whose solution is
\begin{equation}
\label{eq:small_mu_solution_far}
\phi_{\infty}\simeq \frac{c_3 J_{2 l+1}\left(2 \sqrt{r} \mu \right)+c_4 Y_{2 l+1}\left(2 \sqrt{r} \mu \right)}{\mu\sqrt{r}}(1+\OO(1/r))
\end{equation}
where $J$ and $Y$ are Bessel functions.
The reader should appreciate that the near horizon expansion keeps $\mu\ln(r-1)$ finite, while the far region is at finite $\mu\sqrt{r}$. In both cases there is a double scaling limit involving $\mu$ and $r$. Subleading terms are negligible with respect to these finite quantities.
\bigskip

Boundary layer theory also prescribes the (intuitive) matching condition: the near horizon layer is matched to the finite $r$ solution under the assumptions $\mu\ln(r-1)\gg1$, $r-1\ll 1$.
All terms that we can compute in both regions should coincide.

The near horizon and intermediate $r$ expansions are
\begin{align}
e^{-i\mu\ln(r-1)}+\OO(r-1)\simeq1-i\mu\ln(r-1)-\frac{\mu^2}{2}\ln^2(r-1)+\dots+\OO(r-1)\\
\phi\simeq\frac{c_1 4^{-l} \Gamma(l+1)}{\left(\frac{1}{2}\right)_l}-\frac{c_2 \Gamma (2l+2) \log(r-1)}{\Gamma (l+1)^2}+\OO((r-1),\mu^2)
\end{align}
where $(a)_l$ is the Pochhammer symbol. Correctly, all terms either match or are unknown (e.g. $-\frac{1}{2}\mu^2\ln^2(r-1)$ is $\OO(\mu^2)$ at intermediate $r$), provided
\begin{equation}
\label{eq2:c1c2_coefficients}
c_1 = \frac{4^l \left(\frac{1}{2}\right)_l}{\Gamma (l+1)},\quad c_2 = \frac{i \mu \Gamma (l+1)^2}{\Gamma (2 (l+1))}
\end{equation}
\smallskip

Similarly, we now match at large $r$ ($r\gg1$, $\mu^2 r\ll1$).
We obtain the expansion of the scalar in the far region
\begin{equation}
\label{eq2:blayerfar}
(\mu^2r)^l \left(\frac{c_3}{\Gamma (2 l+2)}+c_4(\OO(1))\right)(1+\OO\left(\mu^2r\right))-\frac{c_4 (\mu^2r)^{-l-1} \Gamma (2 l+1)}{\pi }(1+\OO\left(\mu^2r\right))
\end{equation}
and in the intermediate region%
\begin{equation}
\label{eq2:blayerfar-intermediate}
c_1 r^l(1+\OO\left(\frac{1}{r}\right))+\frac{c_2}{r^{l+1}}(1+\OO\left(\frac{1}{r}\right)) + \OO(\mu^2)
\end{equation}
We now match the two by first obtaining some simple estimates: matching terms of order $r^{-l-1}$, given $c_2\sim\mu$, implies $c_4\sim \mu^{2l+3}\ll1$. Considering now $r^l$ we get $c_3\sim\mu^{-2l}\gg1$. The conclusion is that we can safely forget about $c_4$. We finally get
\begin{equation}
c_3 = \mu^{-2l}\Gamma(2l+2) c_1 = \mu^{-2l}\Gamma(2l+2)\frac{4^l \left(\frac{1}{2}\right)_l}{\Gamma (l+1)}
\end{equation}
\medskip

Lastly, expanding the Bessel function at $r\rightarrow\infty$, we obtain
\begin{equation}
\label{eq:small_mass_large_r}
\phi\simeq c_3 \frac{(-1)^l}{\sqrt{\pi}r^{3/4}} \mu^{-3/2}\cos\left(2\mu\sqrt{r}-\frac{3}{4}\pi\right)
\end{equation}
which has an overall $\mu$ scaling of $\mu^{-3/2-2l}$, thus growing with $l$. The physics underlying this result is simple: at higher angular momentum, light particles can rotate around the black hole, thus significantly enhancing the field. Indeed, this scaling can be deduced from a simpler argument: neglecting dimensionless factors, the leading-$r$ behaviour in the intermediate-$r$ regime is $r^l$, which at the matching point $r\sim\mu^{-2}$ amounts to $\mu^{-2l}$. Similarly, $\mu^{-\frac{3}{2}-2l}/r^{3/4}\sim\mu^{-2l}$ at the same matching point.
Due to analytic continuation in $l$, the result also agrees with the one obtained by Hui {\it et al.} for $l=0$.

\subsubsection{Love numbers}
It is worth pointing out that, for intermediate distances, $\phi$ behaves as a massless scalar would, up to subleading $\mu$ corrections. The large $r$ behaviour of $\phi_{\text{intermed}}$ showcased in equation~\ref{eq2:blayerfar-intermediate} suggests, for a massive scalar perturbation with $\mu r_\BH\ll 1$, a definition of the Love numbers mirroring the one for the massless case: $c_2/c_1$ (computed in \cref{eq2:c1c2_coefficients}). Since the result will turn out to be complex, this quantity will be called the scalar response coefficient (its real part being the actual Love number, the imaginary part being related to dissipation), and was calculated for example by Dubovsky et al~\cite{Charalambous:2021mea} (their equation~6.17) for rotating black holes but a strictly massless scalar. Riotto et al.~\cite{Kehagias:2022ndy} were able to reproduce this result as well, using near horizon approximations (see their equation~5.24). We obtain
\begin{equation}
    \label{eq:love_numbers}
    \frac{c_2}{c_1} = \frac{i \mu\, r_\BH\, l!^3}{4^l \left(\frac{1}{2}\right)_l(2l+1)!}
\end{equation}
which is also identical to their formula after some mathematical manipulations and after choosing $\omega=\mu$, as we do throughout. This agreement is not surprising, because the mass $\mu$ of the scalar only affects the solution at large distances (the frequency $\omega$ terms dominates the near horizon zone, while the mass $\mu$ dominates at large distances; the factor $\mu$ in our equations~\ref{eq2:c1c2_coefficients},\ref{eq:love_numbers} would more generally be $\omega$). It makes sense that, in the limit $\mu\rightarrow0$, the massless behaviour is recovered (at finite distances). The are however two nontrivial physical points:
\begin{enumerate}
    \item When $\mu r_\BH\ll1$, there is an intermediate $r$ region where the field behaves as if it were massless, allowing us to \emph{define} the response coefficients and Love numbers. In the opposite regime $\mu r_\BH\gtrsim 1$, we do not have this parametric separation of scales and defining these quantities becomes much harder. We do not attempt to do so in this work.%
    \item Our method, besides its simplicity and generality, allows the computation of subleading corrections to~\ref{eq:love_numbers} in $\omega$ and $\mu$ (even taken to be different).
\end{enumerate}
\medskip

\begin{table}[ht!]
    \centering
    \begin{tabular}{cc}
        \centering
        \begin{tabular}{c}
            \textbf{near horizon}\\
            $1<r\lesssim1+\mu^2$
        \end{tabular}
        & \mymath[colback=gray!10, sharp corners]
        {(r-1)^{\pm i\mu}}
        \\[30pt]

        \begin{tabular}{c}
            \textbf{intermediate} $r$\\
            $1+\mu^2\lesssim r \lesssim \mu^{-2}$
        \end{tabular}
        &  \mymath[colback=gray!10, sharp corners]
        {\frac{4^l \left(\frac{1}{2}\right)_l}{l!} r^l \, _2F_1(-l,-l;-2 l;1/r)+\frac{i \mu (l!)^2}{(2l+1)!}\frac{1}{r^{l+1}} \, _2F_1(l+1,l+1;2 l+2;1/r)}\\[30pt]

        \begin{tabular}{c}
            \textbf{large} $r$\\
            $\mu^{-2}\lesssim r$
        \end{tabular}
        &  \mymath[colback=gray!10, sharp corners]
        {\mu^{-2l-1}(2l+1)!\frac{4^l \left(\frac{1}{2}\right)_l}{l!}\,\frac{J_{2 l+1}\left(2 \mu \sqrt{r} \right)}{\sqrt{r}}}\\[30pt]

        \begin{tabular}{c}
            \textbf{very large} $r$\\
            $\mu^{-2}\ll r$
        \end{tabular}
        &  \mymath[colback=gray!10, sharp corners]
        {(-1)^l(2l+1)!\frac{4^l \left(\frac{1}{2}\right)_l}{\sqrt{\pi}l!}\,\frac{\mu^{-2l-\frac{3}{2}}}{r^{3/4}} \cos\left(2\mu\sqrt{r}-\frac{3}{4}\pi\right)}\\[30pt]
        
    \end{tabular}
    \caption{Behaviour of $\phi$ in the small mass regime.}
    \label{box:small_mass}
\end{table}

We summarize the behaviour of $\phi$ in the three regions in table~\ref{box:small_mass}.
We showcase our results by plotting the real and imaginary part of $\phi$, this time for $\mu=0.1$. 
In order to have smooth transitions from the intermediate region to the boundary layer, we exploit that by assumption both $\phi_{\rm intermed}$ (equation~\ref{intermed}) and $\phi_\infty$ (equation~\ref{eq:small_mu_solution_far}) asymptote to the same $\phi_{\rm b.layer}$ (equation~\ref{eq2:blayerfar}, \ref{eq2:blayerfar-intermediate}) when $r\gg1/\mu^2$, $r\ll1/\mu^2$ respectively.

\begin{figure}
	\centering
	\begin{subfigure}{0.49\textwidth}
		\centering
		\includegraphics[width=1.\linewidth]{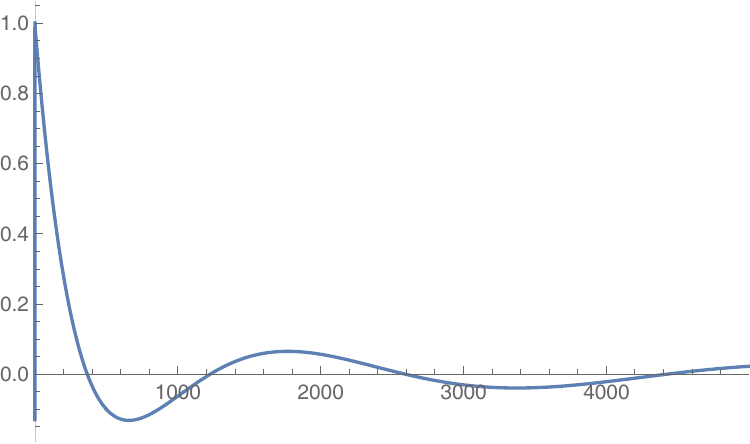}
		\caption{$l=0$. Real part}
	\end{subfigure}
	\begin{subfigure}{0.49\textwidth}
		\centering
		\includegraphics[width=1.\linewidth]{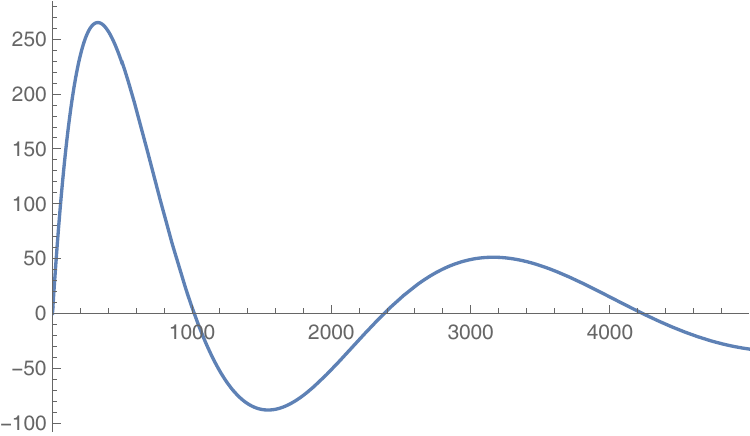}
		\caption{$l=1$. Real part}
	\end{subfigure}
	\centering
	\begin{subfigure}{0.49\textwidth}
		\centering
		\includegraphics[width=1.\linewidth]{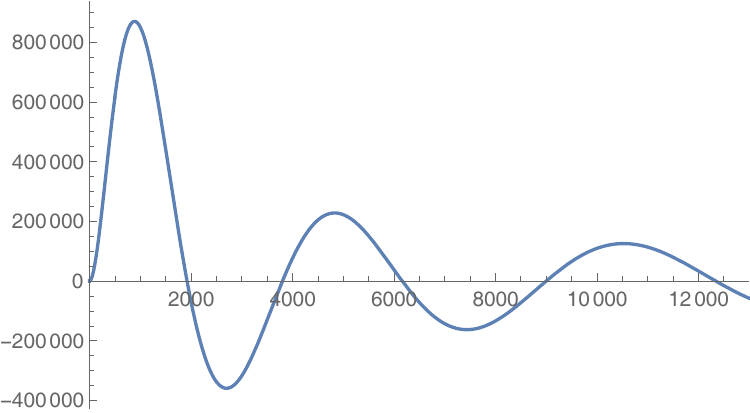}
		\caption{$l=2$. Real part}
	\end{subfigure}
	\begin{subfigure}{0.49\textwidth}
		\centering
		\includegraphics[width=1.\linewidth]{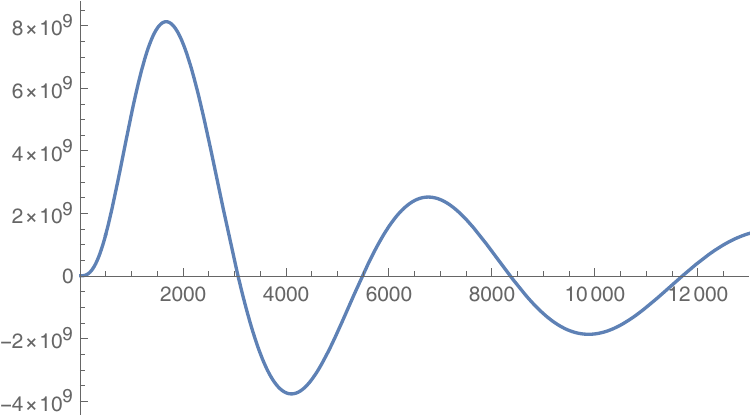}
		\caption{$l=3$. Real part}
	\end{subfigure}
	\caption{Plot of $\phi$ (table~\ref{box:small_mass}) for $\mu r_{\rm{BH}}=0.1$. We normalized the amplitude to $1$ at the horizon. Due to the vast differences in the amplitudes, we plot each $l$ separately.}
	\label{fig:musmall}
\end{figure}

We plot in figure \ref{fig:musmall} the function
\begin{equation}
\phi_{\rm intermed}+\phi_\infty-\phi_{\rm b.layer}
\end{equation}
which has all the correct asymptotic limits. An identical story plays out for the near horizon layer.

The accretion rate for the small mass regime was computed by Hui et al.\ in~\cite{Lam}, who expressed it in terms of the DM density at intermediate distances for $l=0$. Expressing it instead in terms of the density at large distances $r\gg \mu^{-2}$ and dropping $\OO(1)$ factors we get as a function of $l$
\begin{equation}
    \Phi = 8\pi \mu^2 |A|^2,\quad \rho= 2\mu^2|\phi|^2\simeq\mu^2|A|^2\frac{\mu^{-3-4l}}{r^{3/2}}
    \Rightarrow \Phi\simeq \mu^{3+4l}r^{3/2}\rho
\end{equation}
where $A$ is the amplitude of the wave at the horizon $\phi=A(r-1)^{-i\mu}$. The rate is heavily suppressed as $l$ increases, again due to the centrifugal barrier kicking in for small masses.

\bigskip

\subsection{Numerical checks}
We check the above formulas against the numerical results obtained in \cite{Hui:2022sri} by Hui et al.\ In their figure 9, the authors plot the inverse of the amplitude at $r=400r_{\rm{BH}}$ as a function of $\mu$. They normalized the amplitude at the horizon to 1 and took $400r_{\rm{BH}}$ to be far enough that all our far region approximations should apply: indeed the largest length scale that appears in our problem is $1/\mu^2$, which is much smaller than $400$ for $\mu\gg0.05$.
\smallskip

Their plot roughly captures the plateau at large $\mu$, however in the small mass regime the ratio they plot goes like $\sim\frac{1}{\cos(2\mu\sqrt{r})}|_{r=400}$, meaning that one expects large fluctuations when numerically sampling different values of $\mu$. To circumvent this problem, the authors actually average over small intervals in $\mu$. This is a potential drawback of that plot, which ends up depending on the smoothing procedure.
\medskip

Let us now move on to figure 10 of~\cite{Hui:2022sri}, where the authors plot $|\phi|^2$, normalized to 1 at the horizon. They fix $l=m=2$, although $m$ is irrelevant in the spherically symmetric case we are dealing with. The first plot is for $\mu=0.01$, meaning that the profile should accurately be described by our intermediate $r$ region, which extends from $r=1+10^{-4}$ to $10^4$. The analytic expression is
\begin{equation}
    \phi \simeq \frac{4^l \left(\frac{1}{2}\right)_l r^l \, _2F_1\left(-l,-l;-2 l;\frac{1}{r}\right)}{\Gamma (l+1)} \overset{l=2}{=} 1 - 6 r + 6 r^2
\end{equation}
Correctly, their plot presents no oscillations. The order of magnitude of the field is also correctly reproduced by this formula.
\smallskip

Next, let us look at the $\mu=0.5$ case. Here we are always in the large distance regime ($r\gg24$) and we expect our small mass approximation $0.5\ll1$ to give results correct up to $\OO(1)$ corrections. We observe instead that the plot in~\cite{Hui:2022sri} greatly underestimates, by a factor $\sim 10^4$, the field at large distances. To explain this discrepancy, observe that the authors seem to have normalized the field too far from the horizon, missing a sharp drop in the dark matter density close to $r_{\rm BH}$.%
\smallskip

For $\mu=1$ the discussion in appendix~\ref{app:intermediate_mu} implies that intermediate masses behave roughly like large masses, hence our agreement with the decreasing profile.
\medskip

\begin{figure}[t!]
\centering
\begin{subfigure}{0.45\textwidth}
\centering
\includegraphics[width=1.\linewidth]{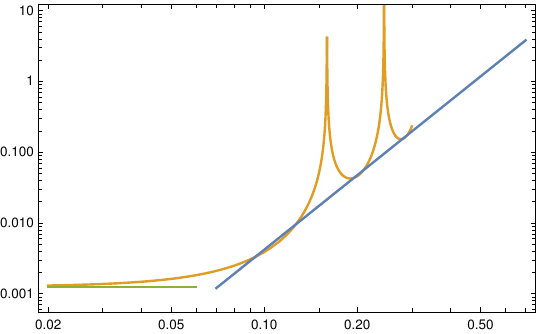}
\end{subfigure}
\begin{subfigure}{0.45\textwidth}
\centering 
\includegraphics[width=1.0\linewidth]{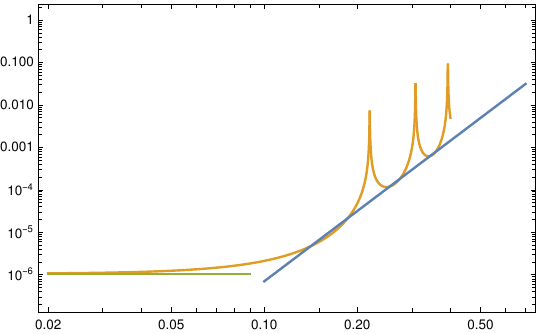}
\end{subfigure}
\caption{Plot of $|\phi(r=1)|/|\phi(r=400)|$ as $\mu$ changes. $l=1$ on the left, $l=2$ on the right. This plot corresponds to figure 11 in~\cite{Hui:2022sri}.
Referring to the equations for $\mu\ll1$ in table~\ref{box:small_mass}, the green line is the intermediate $r$ regime, the orange line is the far regime and the blue line is the very far regime once we set the cosine to $1$, to tame the spikes. The transition between the green and the blue line is smooth if we consider the interpolating orange line.}
\label{fig:Guanhao}
\end{figure}

Finally, let us look at figure 11 in~\cite{Hui:2022sri}, where they plot $1/|\phi|$ at $r=400r_{\rm{BH}}$ for different values of $\mu$. There are roughly three regimes: large mass, small mass and $r\gg\frac{1}{\mu^2}$, small mass and $r\ll\frac{1}{\mu^2}$. The authors provide estimates for all three regimes in their formula (4.1). Our formulas agree except when $r$ is in the far regime $r\gg\frac{1}{\mu^2}\gg1$, which is their intermediate $\mu$ region, where the log-plot linearly grows. In our approximation $|\phi|$ should be given by
\begin{equation}
\label{eq:numericalEnricoLam}
    |\phi(r)|\propto
    \begin{cases}
        r^{-\frac{3}{4}},\quad \mu\gg1\\
        \mu^{-\frac{3}{2}-2l}r^{-\frac{3}{4}},\quad \frac{1}{\sqrt{r}}\ll\mu\ll1\\
        r^{l},\quad \mu\ll\frac{1}{\sqrt{r}}
    \end{cases}
\end{equation}
so the slope $\frac{3}{2}+3l$ in~\cite{Hui:2022sri} should be replaced by $\frac{3}{2}+2l$, not as steep. A simple argument in favour of this is that at $r\sim\mu^{-2}$ the last two formulas in~\ref{eq:numericalEnricoLam} should parametrically coincide. This works well when $\mu$ is small, but the agreement with the numerical solution decreases as the graph plateaus on the right; there the approximation by Hui et al works better. We plot our approximation in figure~\ref{fig:Guanhao}.

The transition to the plateau on the right is predicted to happen when $\mu$ reaches $\mu_c$ given by formula~\ref{eq:critical_mass}, thus the plateau region should move to the right as $l$ increases. Contrary to what we see from the approximations in~\cite{Hui:2022sri}, the transition between the plateau on the left and the region with linear growth in figure~\ref{fig:Guanhao} is smooth once we properly take into account the large $r$ regime (see table~\ref{box:small_mass}) described by a Bessel function.
These examples demonstrate how employing a simple analytic approximation of the solution offers better control over the results compared to the challenges associated with using the full Heun function.

\subsection{Conserved current}
\label{sec:conserved_J}
We pause to introduce a quantity we can prove is conserved on the equations of motion (very closely related to the wronskian). The beauty of the statement is that it is rigorous, exact, and it allows us to prove some claims. The downside is that it will not be enough to solve the problem. We define
\begin{equation}
J\eqdef r(r-1)\left(\phi^*\deriv{\phi}{r}-\phi\deriv{\phi^*}{r}\right)
\end{equation}
A simple substitution reveals
\begin{equation}
\deriv{J}{r} = 0 \quad\text{on-shell}
\end{equation}
Notice that $J$ is conserved as a function of $r$ (not time). Evaluated near the horizon on the causal (i.e. purely infalling) solution, one gets
\begin{equation}
\phi\simeq A (r-1)^{-i\mu},\qquad J = -2i\mu |A|^2
\end{equation}
which, up to a factor, is the energy flux the wave carries into the black hole.
\smallskip

A simple thing we can check is that, if $\omega<\mu$, bounded solutions are not allowed. WKB (which is locally accurate away from turning points) gives $\phi\sim e^{-\sqrt{\mu^2-\omega^2} r}$ at large distances. For this solution, $J$ equals $0$ due to the exponential suppression. Since the causal solution has $J\neq0$, it never matches with a bounded profile for energies lower than the mass. Physically, this translates in the absence of (stable) bound states. We will later devote much discussion to solitons, which are bound states that evade this argument by having complex $\omega$.
\bigskip

We can additionally check whether our approximate solutions respect this conservation law. The answer is trivially yes for the large mass regime approximation. However, in the small mass regime there seems to be a contradiction because $J=0$ at large distances. Is our solution inaccurate?

The answer lies in the size of the ingoing and outgoing wave at large distances: within our accuracy the two amplitudes are large ($\propto \mu^{-3/2-2l}$), which would naively suggest an equally large $J$ at infinity. However because the amplitudes are precisely equal at this order of approximation, the waves sum to a cosine and the leading contribution to $J$ cancels. We expect subleading orders in $\phi$ to lift this degeneracy and give a value of $J$ much smaller than the individual contributions, consistent with the conservation of the current. We conclude that our approximate solution correctly captures the leading order behaviour of the exact solution, passing a nontrivial consistency check.

We will now see that $J$ allows us to compute this slight difference in the amplitudes far away, without the need of computing the aforementioned subleading orders of $\phi$. Very generally,
\begin{align}
\phi_{\rm n.h.}=\phi_{\rm n.h.}^{in}\simeq (r-1)^{-i\mu},\\
\phi_\infty=A_+\phi_\infty^++A_-\phi_\infty^-\simeq A_+ \frac{e^{2i\mu\sqrt{r}}}{r^{3/4}}+A_- \frac{e^{-2i\mu\sqrt{r}}}{r^{3/4}}
\end{align}
where the equal sign is an exact statement, while the $\simeq$ indicates that we are neglecting higher order terms which however do not affect $J$.

Imposing conservation one obtains $|A_-|^2-|A_+|^2=1$. Now we input that, when $\mu\ll1$, $A_-\simeq A_+\simeq A_\infty$, where $A_\infty$ the coefficient in~\ref{eq:small_mass_large_r}. Then simple algebra gives
\begin{equation}
(|A_-|+|A_+|)(|A_-|-|A_+|) = 1\rightarrow 2A_\infty(|A_-|-|A_+|) \simeq 1
\end{equation}
Therefore, as promised, $J$ allows us to compute the very small difference $|A_-|-|A_+|$.

\subsection{Solitons in black hole domination}
\label{sec:soliton_BH_domination}
As discussed in the introduction (see among others~\cite{cardoso_parasitic_2022,bar_looking_2019,bar_galactic_2018,Brax:2019npi}), ultralight dark matter forms solitons.

Dark matter solitons are not topologically protected: we will use the word to loosely mean any stationary field configuration in a bound state, thus the density of dark matter in a compact region will be significantly greater than in the rest of the halo. We will assume the following hierarchy of scales:
\begin{equation}
	r_{\text{Schwarzschild}}\ll r_{\rm{sol}}\ll r_{\text{halo}}
\end{equation}
The scenario we will consider is that the supermassive black hole is at the centre of the soliton. Given this hypothesis,
the behaviour of such solitons can be dominated by the gravity of the supermassive black hole or by the self-gravity of dark matter. We begin the discussion from the first, simpler case.
\smallskip

What is the difference between the discussion of black hole domination of the previous section and the following, which we will call black hole dominated soliton? Self-gravity is still negligible but now we will consider bound states, implying \textquotedblleft $\omega<\mu$ \textquotedblright. Actually $\omega$ will turn out to have a very small imaginary part, evading the argument in section~\ref{sec:conserved_J}. We will first neglect the imaginary part and later compute it and check this assumption.

Denoting $\omega = \mu(1+\gamma)$ and expanding in small $\gamma$, the potential in equation~\ref{eq:eq_diff} now has a term $\propto \gamma r^4$. We will assume this term to be small enough that only the tail of the solutions found in the previous sections are modified, at a distance $r\gtrsim|\gamma|^{-1}$. Seeking only the behaviour of the solution at large distances, we can rewrite~\ref{eq:eq_diff} approximately far from the horizon as
\begin{equation}
    \label{eq:eq_diff_soliton_BH_domination_r}
    r^2\deriv{}{r}\left(r^2\deriv{\phi}{r}\right) +\mu^2\left(2\gamma r^4+r^3\right)\phi=0
\end{equation}
To emphasize the scale $r=|\gamma|^{-1}$, we define $r\equiv \frac{x}{-\gamma}$, obtaining
\begin{equation}
    \label{eq:eq_diff_soliton_BH_domination}
    x^2\deriv{}{x}\left(x^2\deriv{\phi}{x}\right) +\frac{\mu^2}{-\gamma}\left(x^3-2x^4\right)\phi = 0
\end{equation}
This equation can be solved exactly, but a better intuition for the different distance regimes comes from a WKB approximation. To take this route, we now assume $\frac{\mu^2}{|\gamma|}\gg1$. Given $|\gamma|\ll1$, this condition is obvious unless $\mu$ is in the small mass regime. The condition is equivalent to $\frac{1}{|\gamma|}\gg\frac{1}{\mu^2}$, where $\frac{1}{\mu^2}$ is the length scale at which the decaying $r^{-3/4}$ behaviour starts, in agreement with our hypothesis that only the tail of the solution should be affected. We shall further comment on this hypothesis later.

Such a large parameter invites a WKB approximation, which will be accurate far from the turning point $x=\frac{1}{2}$, where the Airy matching will be employed. Then the decaying solution at $x>1/2$ will be
\begin{equation}
    \phi = \frac{A}{(2r^4-r^3)^{1/4}}e^{-\frac{\mu}{\sqrt{|\gamma|}}\int_{1/2}^x \left(2-\frac{1}{x'}\right)^{1/2}\dd x'} \propto
    e^{-\frac{\mu}{\sqrt{|\gamma|}}\sqrt{2}x},\quad x\gg \frac{1}{2}
\end{equation}
and at $x\ll1/2$ it matches
\begin{equation}
    \label{eq2:BH_dominated_soliton}
    \phi = \frac{2A}{(r^3-2r^4)^{1/4}}\cos\left(\frac{\mu}{\sqrt{|\gamma|}}\int_{1/2}^x \left(\frac{1}{x'}-2\right)^{1/2}\dd x'+\frac{\pi}{4}\right) \propto
    \frac{2A}{r^{3/4}}\cos\left(2\mu\sqrt{r}+\phi_0\right)
\end{equation}
We recognize the $r^{-3/4}$ and the oscillating behaviour we already saw in the small mass regime (equation~\ref{eq:small_mass_large_r}). The main observation is that if $\mu r_{\rm{BH}}\gg 1$ there is a surprising clash between causal boundary conditions at the horizon, which demand oscillations that go like $e^{-2i\mu\sqrt{r}}$ (equation~\ref{eq:large_mass_matching}), and boundedness at infinity, which imposes a cosine (equation~\ref{eq2:BH_dominated_soliton}). Black hole dominated solitons can thus only be formed in the small $\mu r_{\rm{BH}}$ regime.

We give the explicit very large distance behaviour of $\phi$, normalizing the amplitude at the horizon to $1$ as usual (that is, matching~\ref{eq2:BH_dominated_soliton} to the very large $r$ regime in table~\ref{box:small_mass})
\begin{empheq}[box={\mymath[colback=gray!10, sharp corners]}]{equation}
    \label{box:BH_dom_soliton_largeR}
    r\gtrsim |\gamma|^{-1},\quad\phi(r) \simeq (-1)^l(2l+1)!\frac{4^l \left(\frac{1}{2}\right)_l}{\sqrt{\pi}l!}\,\frac{\mu^{-2l-\frac{3}{2}}}{2^{5/4}r}e^{-\sqrt{2|\gamma|}\mu r}
\end{empheq}

We now estimate $\rm Im(\gamma)$ and check that it is small. The idea is to estimate black hole accretion, which is linked to the decay of the soliton. Neglecting $\OO(1)$ factors
\begin{equation}
    \deriv{M_\BH}{t} \simeq 4\pi r_\BH^2 \frac{M_\sol}{r_\sol^3}
\end{equation}
so, using $G M_\BH r_\sol \mu^2\sim \OO(1)$,
\begin{equation}
    \rm Im(\gamma) \simeq \frac{1}{M_\sol \mu}\deriv{M_\BH}{t}\sim (\mu r_\BH)^5\ll 1
\end{equation}

Finally we determine whether our hypothesis $(\mu r_\BH)^2\gg|\gamma|$ was well motivated.
Finding $\gamma$ amounts to determining the spectrum of bound states. To this end relying on an exact solution of equation~\ref{eq:eq_diff_soliton_BH_domination} is more appropriate. Demanding boundedness at $x=0$, vanishing boundary condition at $x\rightarrow\infty$ and writing down the full solution in terms of ${}_1F_1$ and $U$ hypergeometric functions, we get the quantization condition
\begin{equation}
    \frac{1}{\Gamma \left(1-\frac{\mu}{2 \sqrt{2|\gamma|}}\right)}=0\rightarrow \frac{(\mu r_{\rm{BH}})^2}{|\gamma|} = 8n^2,\quad n\in\mathbb{N},\,n\ge1
\end{equation}
suggesting that $\gamma$ should indeed always be small.
Had we not known about the exact solution, the quantization condition coming from WKB would have given us $\frac{\mu^2}{|\gamma|}=8(n+\frac{1}{4})^2$. Given this good agreement, in appendix~\ref{app:soliton_frequency} we will trust the WKB result when computing the soliton spectrum in self-gravity domination, which is what we turn to in the next section.

\section{Including self-gravity}
\label{sec:DMsoliton_domination}
A more interesting scenario is instead the soliton dominated regime, meaning that self-gravity dominates over the black hole gravity. We expect this regime to be roughly characterized by the condition $M_{\rm{BH}}\ll M_{\rm{sol}}$, where we define the total soliton mass to be $M_{\sol} = \int_{r_\BH}^\infty \rho_{DM}(r) 4\pi r^2\dd r$ (convergence is guaranteed by the exponential decay at infinity).
Indeed we will see that black hole gravity is not necessary to hold together the soliton: self gravity of dark matter can be enough and we could meaningfully study the case $M_{\rm{BH}}=0$.

One could think of the following possibility: a black hole surrounded by a soliton so massive that the dynamics is entirely dominated by dark matter self-gravity at all distances outside the horizon. We will show that this is actually impossible (assuming stability of the overall system), and that there is always a region outside the horizon dominated by the black hole and general relativity, however small $M_{\rm{BH}}$ may be.

\subsection{Scaling argument for the size of the soliton}
\label{subsec:soliton_size}
We now include the self gravity of the scalar as a (non-relativistic) newtonian potential, which will obey a separate Poisson equation. The aim of the following discussion is to find an estimate for the soliton size. This discussion is fairly standard in the literature, and we will follow~\cite{Hui:2021tkt,bar_galactic_2018}. Once we include self gravity, negative energy solutions appear (bound states). Energy will be minimized by a spherically symmetric field configuration, so we set $l=0$ from here on.

Since it would be confusing to have both the scalar field $\phi$ and the newtonian potential $\Phi$ around, we define $\chi$ as
\begin{equation}
\label{eq:def_chi}
\phi(t,\vec{x}) \eqdef \frac{\chi(r)}{\sqrt{8\pi G}}e^{-i\omega t}
\end{equation}
The $\sqrt{8\pi G}$ factor is chosen to simplify Poisson's equation. Notice how the energy density would be halved if the scalar were real. From now on we will always keep $r_{\rm{BH}}$ explicit.
Equation~\ref{eq:eq_diff} is then amended to
\begin{align}
\label{eq:eq_diff_selfG}
r(r-r_{\rm{BH}})\deriv{}{r}\left(r(r-r_{\rm{BH}})\deriv{\chi}{r}\right)+V(r)\chi=0,\quad
V(r) \eqdef \mu^2\left(-2r^4\Phi+\left[(\gamma+1)^2-1\right]r^4+r^3r_{\rm{BH}}\right)\\
\label{eq:eq_diff_Poisson}
\partial_r^2(r\Phi)-\mu^2 r |\chi|^2=0
\end{align}
where by assumption $\Phi$ vanishes at large distances and $\omega=\mu(1+\gamma)$.
We emphasize that this equation retains a fully relativistic treatment of the black hole in terms of its background metric, on top of which a correction is introduced to account for the self-gravity of the scalar. For a non-relativistic stress-energy tensor, this metric correction is suppressed by powers of $c$ except for the time-time component, which is identified as the Newtonian potential. We further assume that $\Phi$ only affects the dark matter profile far from the black hole, so that the background metric in Poisson's equation can be approximated as flat. We will later observe the self-consistency of this assumption.
At large $r$ equation~\ref{eq:eq_diff_selfG} reduces to
\begin{equation}
\label{eq:eq_diff_selfG_far}
\partial_r^2\left(r\chi\right)+\mu^2(r_{\rm{BH}}-2r\Phi+r\left[(\gamma+1)^2-1\right])\chi=0
\end{equation}
which is the form used in~\cite{bar_galactic_2018}.
\smallskip

In the approximation where the black hole is negligible, a well known scaling symmetry allows one to set $\chi(0)=1$. Then, numerical simulations~\cite{Chavanis:2011zi,Chavanis:2011zm,Marsh:2015wka} indicate that the smallest possible $\gamma$ is $\simeq-0.69$ which partially supports the approximation $|\gamma|\ll1$. Scaled solutions turn out to have larger, less massive solitons and smaller $\gamma$.
\smallskip

To obtain the soliton size, we argue that $GM_{\rm{sol}}\sim G\rho_s r^3\sim\mu^2\chi^2r^3$, while from~\ref{eq:eq_diff_Poisson},\ref{eq:eq_diff_selfG_far} we get $\Phi\sim\mu^2r^2\chi^2,\,\chi\sim\mu^2r^2\Phi\rchi$. Substitutions yield $GM_{\rm{sol}} r_{\rm{sol}}\mu^2\sim1$, which can be compared with the analogous result obtained in black hole domination
\begin{empheq}[box={\mymath[colback=gray!10, sharp corners]}]{equation}
    \label{eq:soliton_size}
M_{\rm{sol}}\gg M_{\rm{BH}}:\, r_{\rm{sol}} GM_{\rm{sol}}\sim \frac{1}{\mu^2},\qquad
M_{\rm{sol}}\ll M_{\rm{BH}}:\, r_{\rm{sol}} GM_{\rm{BH}}\sim \frac{1}{\mu^2}
\end{empheq}

A much more detailed discussion of soliton domination can be found in appendix~\ref{app:soliton_frequency}.
We notice that for $M_{\rm{sol}}\simeq M_{\rm{BH}}$ the two formulas for $r_{\rm{sol}}$ parametrically coincide.

\subsection{Exact results}
\label{subsec:exact_results_selfgravity}

The purpose of this section is twofold. The first will be to exploit some mathematical results already available in the literature and apply them to a simplified version of our problem. Our aim will be to re-derive the same results using our WKB approximation, which will give us confidence in the results we will later find for the full problem under the same approximations. Secondly, we will derive an exact result (essentially the energy balance equation) which we will use to rigorously show that $\omega$ must be complex for bound states. This implies that bound states of the scalar will be slowly eaten away by the black hole, so they can only be meta-stable.

Therefore, we will now adapt some exact results and mathematical theorems, derived by Tod and Moroz in~\cite{Tod_1999}, to our problem. They will be far from enough to solve it, but they will enable us to put some of the our findings on firmer grounds (e.g. the quantization of the soliton spectrum). We discuss the simplified problem of a real scalar and no black hole.
\bigskip

As a first step, we map the system of differential equations~\ref{eq:eq_diff_selfG},\ref{eq:eq_diff_Poisson} onto a problem studied by Tod and Moroz in~\cite{Tod_1999}.
They consider a system of differential equations for what they call $S$ and $V$, which map to our variables through
\begin{equation}
V = -\mu^2\left(2\Phi -\frac{r_{\rm{BH}}}{r}+\left[1-(1+\gamma)^2\right]\right),\qquad
S = \sqrt{2}\mu^2 \chi
\end{equation}
They assume $S\in\mathbb{R}$ (while we work with a complex field) and $S(r)$ bounded and vanishing at infinity. They also assume $r_{\rm{BH}}=0$, so there is no black hole. Their results can be summarized as follows:
\begin{itemize}
	\item There is a discrete family of solutions labelled by $n\in\mathbb{N}$, the $n-th$ solution having $n-1$ zeros and vanishing at large $r$ as $\frac{1}{r}e^{-\mu\sqrt{1-(1+\gamma)^2}r}$
	\item $\gamma_n<0$ and increasing with $n$ towards zero
	\item There are no other solutions
\end{itemize}
To develop a better understanding of Tod and Moroz's findings, we begin by rederiving their results using our WKB approximation.
\medskip

The WKB solution of~\ref{eq:eq_diff_selfG_far} can be immediately written.
Defining $C^2=1-(1+\gamma)^2$, we have
\begin{equation}
\label{eq:wkb_complete_solution}
\chi = \frac{A}{r} \frac{e^{i\int^r\mu\sqrt{-2\Phi-C^2}\dd r}}{(-2\Phi-C^2)^{1/4}}+\frac{B}{r} \frac{e^{-i\int^r\mu\sqrt{-2\Phi-C^2}\dd r}}{(-2\Phi-C^2)^{1/4}}
\end{equation}
for $A,B$ constants. We still don't know $\Phi$, but supposing $-2\Phi-C^2>0$ at large distances and recalling that $\Phi$ vanishes at infinity, we would get $\chi\sim 1/r$ and the total mass would diverge. Thus $-C^2<0$ and $\gamma<0$.

Thus $-2\Phi-C^2<0$ at large $r$, and imposing boundedness we get
\begin{equation}
\chi = \frac{A'}{r} \frac{e^{-\int^r\mu\sqrt{2\Phi+C^2}\dd r}}{(2\Phi+C^2)^{1/4}}
\end{equation}
which has the predicted asymptotic behaviour. As we decrease $r$, we can expect $2\Phi+C^2$ to change sign at some finite distance that we interpret as the soliton size $r_{\rm{sol}}$.

The solution at $r<r_{\rm{sol}}$ will transition from exponentially decaying to oscillating and the coefficients $A,B$ will (in principle) be calculable given $A'$ and $\gamma$ thanks to the WKB connection formulas.

Generically~\ref{eq:wkb_complete_solution} will diverge at the origin, unless $A=-B$. This condition quantizes the frequency spectrum, giving rise to the set of $\gamma_n$. As we increase $\gamma$, $C^2$ decreases approaching zero, while $r_{\rm{sol}}$ increases. In analogy with what happens in quantum mechanics, the number of zeros of the solution increases by one for each new bound state.

This concludes the derivation of the results in~\cite{Tod_1999} using WKB. Notice how we chose $A'$ to be real. We can however rotate it by an overall phase, thus generating {\it all} bounded solutions of the {\it complex} scalar case with a simple overall phase rotation of $\chi$. Having overcome this limitation, we now turn to the second one, namely the lack of a black hole which we now introduce back.
\bigskip

We now reconsider equations~\ref{eq:eq_diff_selfG} and~\ref{eq:eq_diff_Poisson}. They were obtained by treating the black hole within general relativity, and adding the scalar self gravity as a weak newtonian potential. We will now write down the equation for the scalar field and generalize the argument given in section~\ref{sec:conserved_J} to prove that bound states are metastable: their frequency $\omega$ cannot be purely real. We do this by generalizing the previously conserved current: the main result is that $J$ is not conserved anymore, and the deviation from $\deriv{J}{r}=0$ is proportional to the decay rate of the soliton.
\medskip

Assuming spherical symmetry we can choose $g_{\theta\theta}=r^2,\,g_{\phi\phi}=r^2\sin^2\theta$ as usual. We expect the scalar field $\phi$ to deform the Schwarzschild metric and lead generically to some $g_{tt} = -A(r),\,g_{rr} = B(r)$. The equation for the scalar field is
\begin{equation}
\label{eq:equazione_full}
\omega^2r^4\phi+\sqrt{-\frac{A}{B}}r^2\deriv{}{r}\left(\sqrt{-\frac{A}{B}}r^2\deriv{\phi}{r}\right)+Ar^4\mu^2\phi=0
\end{equation}
and $J$ can be straightforwardly defined and shown not to be conserved
\begin{equation}
\label{eq3:J_conservation}
J = \sqrt{-\frac{A}{B}} r^2 \left(\phi^*\deriv{\phi}{r}-c.c.\right),\qquad \sqrt{-\frac{A}{B}}r^2\deriv{J}{r} = -2i\,\text{Im}(\omega^2)\,r^4 |\phi|^2
\end{equation}
We now want to show that $\omega$ cannot be real for soliton configurations. From equation~\ref{eq3:J_conservation} real $\omega$ implies conservation of $J$; however a direct computation reveals that bound states have $J=0$ at infinity due to the exponential suppression of $\phi$, while $\phi$ close to the horizon is still a purely infalling wave, hence $J\neq0$.
Since this violates $\deriv{J}{r}=0$, we have proved our claim $\rm{Im}(\omega)\neq0$.

This fact is sometimes overlooked in the literature, however it has a clear physical counterpart: bound states of the scalar will be slowly eaten away by the black hole, so they cannot be stable. Actually, $\rm Im(\omega)$ is the inverse time in which the bound state decays.
\medskip

We leave it for appendix~\ref{sec:appendix_wavepackets} to present and solve a concern one can have over the way we impose causality at the horizon. When the frequency is complex, $\phi_{\text{causal}}$ diverges at the horizon while $\phi_{\text{non-causal}}$ goes to zero. Since the non-causal solution is already exponentially suppressed at the horizon (and strictly zero at $r=r_{\rm{BH}}$), one might worry that the causality condition has to be rediscussed. Actually, a more careful analysis using wave packets shows that the physical solution is always bounded and the usual causality condition should be imposed. This is reminiscent of black hole quasinormal modes, whose solutions are also divergent in frequency domain.

\subsection{Solution from large to small $r$}
\label{upperbounds}
We now explore the solution of~\ref{eq:eq_diff_selfG_far} as we vary $r$. If the black hole dominates at all distances, we can neglect $\Phi$ when determining $\chi$. As a second step, one can then determine $\Phi$ by solving~\ref{eq:eq_diff_Poisson}. Since $\Phi$ does not affect the scalar profile in this scenario the problem is reduced to the one solved in section~\ref{sec:no_self_gravity}.

In this section we instead assume that the soliton dominates the dynamics at large distances, meaning $M_{\rm{BH}}\ll M_{\rm{sol}}$. One goal will be to estimate $\Phi$ and consequently understand if the assumption $r_{\rm{BH}}\ll|r\Phi|$ breaks down. It turns out that it always breaks down, at a distance we will name $r_{\rm{e}}$. At shorter distances we will then match the solution to the one where we assume $r_{\rm{BH}}\gg|r\Phi|$, which is the regime discussed in section~\ref{sec:no_self_gravity}. Finally, we assume $\rm{Im}(\omega)$ to be negligible compared to $\mu$, since its inverse is the time scale of the soliton decay, which we assume to be long-lived. We will check the self-consistency of this claim once we obtain the decay time of the soliton.
\bigskip

We start with the WKB solution of~\ref{eq:eq_diff_selfG_far} for $\chi$. Since the soliton dominates, $M_{\rm{sol}}\gg M_{\rm{BH}}$ and the results obtained by Tod and Moroz should approximately apply. The WKB approximation was able to reproduce all their results (see subsection~\ref{subsec:exact_results_selfgravity}), so we will trust this solution until the condition $r_{\rm{BH}}\ll|r\Phi|$ is no longer met.
\smallskip

Recall that we labeled the size of the soliton $r_{\rm{sol}}$, which in WKB language coincides with the (single) zero of $2\Phi+C^2$.
The hierarchy of scales (which will be proved in the following) can be summarized as
\begin{equation}
\label{eq:hierarchy}
r_{\rm{BH}}\ll r_{\rm{e}}\ll r_{\rm{sol}}
\end{equation}
We can write the WKB ansatz for the solution both at $r>r_{\rm{sol}}$ where it decays and at $r<r_{\rm{sol}}$ where it oscillates. We then match them using the well known WKB connection formulas, resulting in
\begin{equation}
\label{eq:chi_large_distance}
\chi = \frac{A'}{r} \frac{e^{-\mu\int^r_{r_{\rm{sol}}}\sqrt{2\Phi+C^2}\dd r}}{(2\Phi+C^2)^{1/4}},\;r>r_{\rm{sol}}.\quad \chi = \frac{2A'}{r} \frac{\cos(\mu\int^r_{r_{\rm{sol}}}\sqrt{-2\Phi-C^2}\dd r+\frac{\pi}{4})}{(-2\Phi-C^2)^{1/4}},\;r<r_{\rm{sol}}
\end{equation}
where $C^2=1-(1+\gamma)^2$. We take this to be accurate away from the turning point, where the denominator $(-2\Phi-C^2)$ becomes zero and WKB breaks down while the true solution remains of the same order of magnitude.

Given our experience from the case without a black hole, if we consider the $n$-th bound state, the cosine oscillates $n$ times. Besides, to restrict to physically meaningful solutions we should impose $\cos=0$ at the origin, so that $\chi$ remains finite.
To restrict to the ground state, we should therefore be able to write
\begin{equation}
    \chi = \frac{2A'}{r} \frac{\sin(\mu\int^r_{0}\sqrt{-2\Phi-C^2}\dd r)}{(-2\Phi-C^2)^{1/4}}
\end{equation}
with the understanding that $\mu\int^{r_{\rm{sol}}}_{0}\sqrt{-2\Phi-C^2}\dd r=\frac{3}{4}\pi$

We now make an important simplifying assumption (we will check its validity later): we assume $-2\Phi-C^2$ to always be of the same order $k$ when $r_{\rm{e}}\ll r\ll r_{\rm{sol}}$, so that $\chi\propto\frac{\sin(\mu k r)}{r}$ and we can compute $\rho(r)$ as
\begin{equation}
\label{eq:rho_missing}
4\pi\braket{\rho} \simeq \left\{\begin{array}{ll}
(1.5)\dfrac{M_{\rm{sol}} \sin^2(\frac{3\pi r}{4 r_{\rm{sol}}})}{r^2 r_{\rm{sol}}}, & \text{for } r_{\rm{e}}\lesssim r\lesssim r_{\rm{sol}} \\
(0.76) \dfrac{M_{\rm{sol}}}{r^2 r_{\rm{sol}}}e^{-\frac{3\pi (r-r_{\rm{sol}})}{2 r_{\rm{sol}}}}, & \text{for } r\gtrsim r_{\rm{sol}}
\end{array}\right.
\end{equation}
where $\OO(1)$ factors (not essential) are determined from continuity and imposing that the total mass should indeed be $M_{\rm{sol}}$. The region $r\lesssim r_{\rm{e}}$ will be discussed later. Our approximation neglects the halo which is hosting the soliton. As a consequence, we are unable to discuss the relation between the halo and the soliton mass found in~\cite{Schive:2014dra,Schive:2014hza,Marsh:2015wka}. However we expect the halo to have a negligible effect on the solution, as long as the soliton dominates over the halo density.
\smallskip

We are now in the position to carry out the program outlined at the beginning of this subsection: computing $\Phi$ and estimating $r_{\rm{e}}$.

From Poisson's equation, $\Phi$ can be computed starting from infinity and imposing the boundary condition $\Phi(r)\sim-\frac{GM_{\rm{sol}}}{r}$. It turns out that, as expected, the potential is well approximated by this expression everywhere outside the soliton, despite the exponential tail of the density. Inside the soliton
\begin{empheq}[box={\mymath[colback=gray!10, sharp corners]}]{equation}
\label{eq:Phi}
\Phi(r) \simeq \left\{\begin{array}{ll}
-c_1\frac{G M_{\rm{sol}}}{r}+c_2\frac{G M_{\rm{sol}}}{r_{\rm{sol}}}+(1.5)\left(-\frac{G M_{\rm{sol}} \mathrm{CosInt}\left(\frac{3 \pi  r}{2 r_{\rm{sol}}}\right)}{2 r_{\rm{sol}}}+\right.&\\
\; \qquad\qquad\left.+\frac{G M_{\rm{sol}} \log \left(\frac{r}{r_{\rm{sol}}}\right)}{2 r_{\rm{sol}}}+\frac{G M_{\rm{sol}} \sin \left(\frac{3 \pi  r}{2 r_{\rm{sol}}}\right)}{3 \pi  r}\right), & \text{for } r_{\rm{e}}\lesssim r\lesssim r_{\rm{sol}} \\[15pt]
-\frac{GM_{\rm{sol}}}{r}, & \text{for } r\gtrsim r_{\rm{sol}}
\end{array}\right.
\end{empheq}
We now determine $c_1,c_2$. Not only does $c_1=0$ make $\Phi'(r_{\rm{sol}})$ continuous (up to $8\%$ error), but it makes $\Phi$ finite at the origin as we expect. $c_2\simeq-1$ can be deduced from the continuity of $\Phi(r_{\rm{sol}})$.

A more drastic approximation would have been to average $\chi^2\sim\frac{\cos^2}{r^2}\sim\frac{1/2}{r^2}$, thus giving a potential which goes like $\Phi\simeq \frac{GM_{\rm{sol}}}{r_{\rm{sol}}}(\ln(r/r_{\rm{sol}})-1)$ inside the soliton. This is in good agreement with the more complicated potential we got in equation~\ref{eq:Phi}, except close to the origin, where only the more accurate approximation shows that $2\Phi+C^2$ changes slowly and replacing it with a constant should yield reliable results.
\medskip

We now determine the length scale $r_{\rm{e}}$, where $-2r\Phi\simeq r_{\rm{BH}}$. Assuming $M_{\rm{BH}}/M_{\rm{sol}}\ll 1$ we expect this scale to be much smaller than $r_{\rm{sol}}$. Exploiting this, we can expand $r\Phi$ close to the origin where it is very closely linear. We obtain
\begin{empheq}[box={\mymath[colback=gray!10, sharp corners]}]{equation}
    r_{\rm{e}} \simeq \frac{M_{\rm{BH}}}{(2.3) M_{\rm{sol}}} r_{\rm{sol}}
\end{empheq}
Observe first that $r_{\rm{e}}\ll r_{\rm{sol}}$ and secondly that the black hole dominated region $r_{\rm{BH}}<r<r_{\rm{e}}$ always exists, because stability of the soliton forces $GM_{\rm{sol}}\ll r_{\rm{sol}}$ which implies
\begin{equation}
    \frac{r_{\rm{sol}}}{GM_{\rm{sol}}}\gg1\rightarrow r_{\rm{e}} = r_{\rm{sol}} \frac{M_{\rm{BH}}}{M_{\rm{sol}}} \gg r_{\rm{BH}}
\end{equation}
and therefore $r_{\rm{BH}}\ll r_{\rm{e}}$. This confirms the whole hierarchy of scales we assumed in~\ref{eq:hierarchy}.
\medskip

Is the stability requirement $GM_{\rm{sol}}\ll r_{\rm{sol}}$ trivially satisfied? As discussed in the review~\cite{Hui:2021tkt}, it is not. In section~\ref{subsec:soliton_size} we linked the soliton size to the masses $\mu,M_{\rm{sol}}$. This constrains the possible values of $\mu$ as follows: 
\begin{equation}
\label{eq:upper_bound_soliton}
r_{\rm{sol}} \gg G M_{\rm{sol}}\rightarrow GM_{\rm{sol}} r_{\rm{sol}} \gg (GM_{\rm{sol}})^2\rightarrow \frac{1}{\mu^2}\gg (GM_{\rm{sol}})^2\rightarrow \mu r_{\rm{BH}} \ll \frac{M_{\rm{BH}}}{M_{\rm{sol}}}
\end{equation}
which therefore implies
\begin{equation}
\mu r_{\rm{BH}} \ll 1 \; .
\end{equation}

This is a very general bound that links the microscopic quantity $\mu$ with the masses of astrophysical objects. 
If a soliton is observed in a galaxy that hosts a SMBH with mass $M_{\rm BH}$, it implies an upper bound on the mass of the scalar:
\begin{equation}
\mu \ll 10^{-19} {\rm eV} \Big( \frac{10^9 M_{\astrosun}}{M_{\rm BH}} \Big) \; .
\end{equation}

For completeness, let us point out that solitons dominated by a black hole also have a stability bound $\mu r_{\rm{BH}} \ll \sqrt{\frac{M_{\rm{BH}}}{M_{\rm{sol}}}}$, which however is trivially satisfied because, due to the discussion in~\ref{sec:soliton_BH_domination}, $\mu r_{\rm{BH}}\ll 1$ in this regime.
A novelty that we bring is that the bound for soliton domination also forces us to restrict to the small $\mu r_{\rm{BH}}$ regime, where we know that boundary conditions imposed at the horizon become less and less important the further away we are from $r_{\rm{BH}}$ (in fact the part of the solution that depends on them decays like $\sim1/r$ compared to the part which does not). Given this and $r_{\rm{e}}\gg r_{\rm{BH}}$, we deduce that boundary conditions are unimportant in the regime of soliton domination. Quantitatively, suppression is $\propto\frac{r_{\rm{BH}}}{r_{\rm{e}}}=\frac{GM_{\rm{sol}}}{r_{\rm{sol}}}$.

\subsection{Solitons at small $r$}
We now match the solution obtained thus far with what we got in section~\ref{sec:no_self_gravity} under the approximation that the black hole dominates the dynamics, which happens when $r<r_{\rm{e}}$. From the bound $\mu r_{\rm{BH}}\ll \frac{M_{\rm{BH}}}{M_{\rm{sol}}}\ll1$ we can focus on the small mass regime only.
\medskip

Far from the horizon the field can be approximated as
\begin{equation}
\label{eq:near_horizon_soliton}
\chi \simeq \frac{J_1(2\mu\sqrt{r_{\rm{BH}} r})}{\mu\sqrt{r_{\rm{BH}} r}}
\end{equation}
We now show that $\mu \sqrt{r_{\rm{BH}} r_{\rm{e}}}\ll1$, meaning that we are close to maximum of $\chi$. In the language of section~\ref{sec:no_self_gravity}, this is the intermediate $r$ region.
Some simple algebra and the definitions of $r_{\rm{e}}$ and $r_{\rm{sol}}$ lead to
\begin{equation}
\left(\frac{r_{\rm{BH}}}{GM_{\rm{sol}}}\right)^2\ll1 \rightarrow \mu^2 r_{\rm{sol}} \frac{r_{\rm{BH}}^2}{GM_{\rm{sol}}}\ll 1 \rightarrow \mu\sqrt{r_{\rm{BH}} r_{\rm{e}}} \ll 1
\end{equation}
We can then approximate $\chi$ with a constant. The density is matched at $r=r_{\rm{e}}$,
\begin{equation}
\rho(r) = \rho(r_{\rm{e}}) = (8.3)\frac{M_{\rm{sol}}}{4\pi r_{\rm{sol}}^3}
\end{equation}
\ref{eq:near_horizon_soliton} was derived assuming $\gamma=0$, but this doesn't matter. In fact, going through the derivation of the wave profile one has to replace $\mu\rightarrow\omega$ in the expression of the wave near the horizon, which gives a negligible contribution if $r\ll \frac{1}{\mu^2 r_{\rm{BH}}}$. %
Up to very small corrections $\propto \frac{M_{\rm{BH}}^2}{M_{\rm{sol}}^2}$, this density is what one would obtain had they neglected the black hole altogether, which is the usual assumption in the literature. The point of our analysis was to check that the part of the solution dependent on the boundary conditions never enters a nonlinear regime.
\medskip

We can thus fill in the missing piece of equation~\ref{eq:rho_missing}:
\begin{empheq}[box={\mymath[colback=gray!10, sharp corners]}]{equation}
\label{eq:rho_full}
4\pi\braket{\rho} \simeq \left\{\begin{array}{ll}
(8.3) \dfrac{M_{\rm{sol}}}{r_{\rm{sol}}^3}, & \text{for } r\rightarrow 0 \\
(1.5)\dfrac{M_{\rm{sol}} \sin^2(\frac{3\pi r}{4 r_{\rm{sol}}})}{r^2 r_{\rm{sol}}}, & \text{for } r\lesssim r_{\rm{sol}} \\
(0.76) \dfrac{M_{\rm{sol}}}{r^2 r_{\rm{sol}}}e^{-\frac{3\pi (r-r_{\rm{sol}})}{2 r_{\rm{sol}}}}, & \text{for } r\gtrsim r_{\rm{sol}}
\end{array}\right.
\end{empheq}

\begin{figure}[t!]
\centering

\begin{subfigure}{0.45\textwidth}
\centering
\includegraphics[width=1.\linewidth]{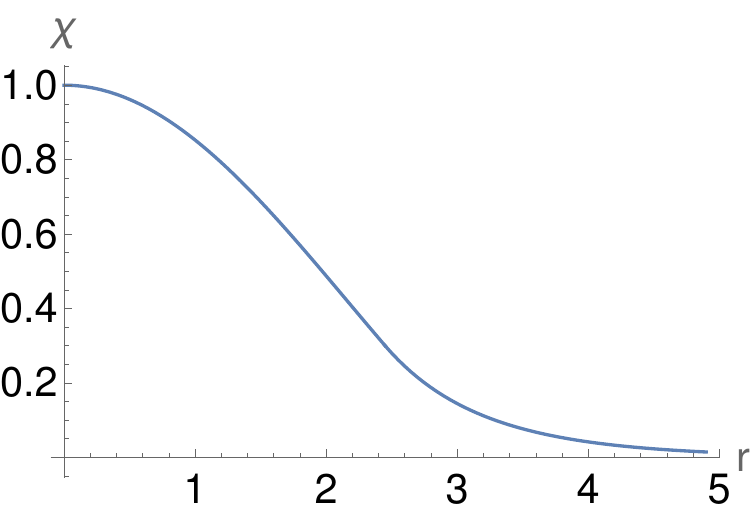}
\end{subfigure}
\begin{subfigure}{0.45\textwidth}
\centering 
\includegraphics[width=1.0\linewidth]{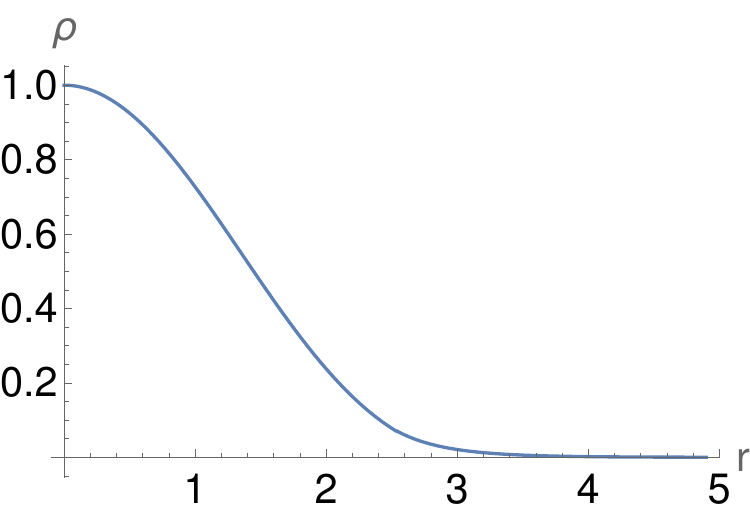}
\end{subfigure}
\caption{Analytic soliton profile for $r_{\rm{sol}}=2.45,\,M_{\rm{sol}}=1.8$.}
\label{fig:soliton}
\end{figure}

We present the plot of this function in figure~\ref{fig:soliton}.
\medskip

We end this section computing the accretion rate of the black hole. This quantity depends on the precise value of $\rho$ at the horizon. Matching $\rho_{hor}$ to the soliton profile outside the sphere of influence of the black hole was always achieved under some simplifying assumptions (i.e. absence of features in the dark matter density profile inside the black hole dominated region) we now have under control.
Employing~\ref{eq:rho_full} we have
\begin{equation}
    \label{eq:accretion1}
    \deriv{M_{\rm{BH}}}{t} = 4\pi r_{\rm{BH}}^2 \rho_{hor}\simeq (8.3) r_{\rm{BH}}^2 \frac{M_{\rm{sol}}}{r_{\rm{sol}}^3}
\end{equation}
We can rewrite $r_{\rm{sol}}$ using $GM_{\rm{sol}}r_{\rm{sol}}\mu^2\simeq\OO(1)$, where the numerical constant is computed in appendix~\ref{app:soliton_frequency} to be $\simeq 6$ (note that some authors report roughly half). Then we have
\begin{equation}
    \label{eq:accretion2}
    \deriv{M_{\rm{BH}}}{t} = (0.04) r_{\rm{BH}}^2 G^3 M_{\rm{sol}}^4 \mu^6
\end{equation}
which agrees with the numerical results in equation (11) in ref.~\cite{cardoso_parasitic_2022} (their numerical prefactor corresponds to $0.06$).

As promised at the beginning of section~\ref{upperbounds}, we now check that $\rm{Im}(\omega)\ll\mu$. From~\ref{eq:accretion1} we can extract
\begin{equation}
    \rm{Im}(\gamma)= \frac{\rm{Im}(\omega)}{\mu}\simeq \frac{\frac{\dd M_\sol}{\dd t}}{\mu M_\sol} \propto \frac{r_\BH^2}{\mu r_\sol^3}
\end{equation}
To see that this is $\ll1$, we square and note the following inequalities
\begin{equation}
    \frac{r_{\BH}^4}{\mu^2r_{\sol}^6}\sim \frac{r_{\BH}^4}{r_{\sol}^6}GM_{\sol}r_{\sol}\ll\left(\frac{r_\BH}{r_\sol}\right)^4\ll1
\end{equation}
from the stability of the soliton.

\section{Discussion}
We reviewed black holes endowed with nonrotating wave dark matter hair and extended the analysis to $l>0$. When $\mu r_{\rm{BH}}\ll1$ interference patterns typical of wave dark matter emerge, giving rise to $\OO(1)$ fluctuations in the density profile. However, in the opposite regime, these effects are washed out by the black hole gravity. This last conclusion crucially relies on imposing causal boundary conditions at the horizon, suggesting that caution should be used when modelling a black hole using newtonian gravity, even at large distances. In addition to providing control over all distances, our techniques offer a precise method for calculating the dynamical Love numbers for scalar perturbations to leading order in small $\omega$ and $\mu$, which can be easily extended to subleading orders.

Our primary focus is on dark matter solitons centered around a supermassive black hole. In a black hole dominated soliton, $M_{\rm{BH}}\gg M_{\rm{sol}}$, the scalar profile is given by the solution presented by Hui et al.\ up to distances $r\sim r_{\rm BH}/|\gamma|$ beyond which it exponentially decays. As we discuss in section \ref{sec:soliton_BH_domination}, such solution exists only in the small mass regime $\mu r_{\rm BH} \ll 1$. Solutions with larger mass are incompatible with causal boundary condition at the black hole horizon.

In the opposite regime, $M_{\rm{BH}}\ll M_{\rm{sol}}$, we analytically computed the approximate soliton profile at large (small) distances where gravity is dominated by the soliton (black hole), respectively.  At the radius $r_e$, representing the sphere of influence of the black hole, we matched the density profiles. The soliton's stability in this scenario enforces a hierarchy of scales, $r_{\rm{BH}}\ll r_e\ll r_{\rm{sol}}$, which also implies a small mass limit, $\mu r_{\rm BH} \ll M_{\rm{BH}}/M_{\rm{sol}}$. This condition indicates that causal boundary conditions at the horizon have no effect far from $\rm{BH}$, similar to the small mass regime in black hole domination.

We emphasize that having analytical control over the solution enables us to perform these assessments. With a solution that remains valid even at horizon scales, we can reliably compute the accretion rate of the black hole.
\smallskip

We now address some limitations of our work. Firstly, we do not model the dark matter halo, which is a notoriously challenging task. However, we expect that our results remain applicable as long as the dark matter density within the soliton overwhelms the halo density. Performing a matching of our solution with the halo profile is beyond the scope of this work.

Secondly, we have focused on non-rotating black holes. We do not believe this assumption to be overly restrictive, given that Sagittarius is a slowly spinning black hole with $J/M<10\%$.  In addition~\cite{Hui:2022sri} numerically studied the issue and found moderate corrections for $J/M$ up to $50\%$. Nevertheless, accounting for a (slowly) spinning black hole would provide an interesting cross-check of those results. Because the analogue of equation~\ref{eq:eq_diff} for spinning black holes can be similarly written, we expect the analysis to be feasible using techniques akin to the ones applied in this paper. However, we leave this investigation for future research.

Lastly we neglected any non-gravitational interaction. A discussion of the validity of this assumption can be found for example in~\cite{bar_galactic_2018}, where the authors conclude that a wide range of ALPs can satisfy this approximation.

\section{Aknowledgments}

We thank Lam Hui, Luca Santoni, Guanhao Sun, Giovanni Maria Tomaselli, and Rodrigo Vicente for helpful discussions and comments on the draft.

\appendix
\section{Intermediate mass $\mu$ at $l\ge1$}
\label{app:intermediate_mu}
This appendix is dedicated to developing an approximate solution of~\ref{eq:eq_diff} for the white are in figure~\ref{fig:validityofapprox}.

The small $\mu$ approximation is by definition invalid, so we would like to again resort to WKB. The main complication is that now $V(r)$ has two zeros. To tackle this problem, we fully exploit the technique of uniform approximations we introduced in section~\ref{sec:no_self_gravity}.

We will first find a differential equation (that we can explicitly solve) with the same structure of zeros as~\ref{eq:eq_diff} in the region $r>1$. We will then map this exact solution to an approximate solution of the original~\ref{eq:eq_diff}.
\bigskip

We keep the definitions of $\tilde{r},\tilde{x}$ given in~\ref{eq:def_rtilde},\ref{eq:def_xtilde}. We now also have the zeros of $V(r)$, given by $r_{1,2}$ as defined in~\ref{eq:def_r12}.

As auxiliary differential equation, we choose~\ref{eq:eq_diff_wkb} where $\Gamma$ should now capture the zeros of $V(r)$. The simplest choice is $\Gamma(x)\eqdef k(x-r_1)(x-r_2)$, $k>0$. The zeros will truly match only if we impose (see~\cite{Berry} for a detailed discussion of this point)
\begin{equation}
\label{eq:def_uniform_k}
\int_{r_1}^{r_2}\sqrt{V(r)}\dd\r \equiv \int_{r_1}^{r_2}\sqrt{\Gamma}\dd\x\rightarrow \int_{r_1}^{r_2}\frac{\sqrt{V(r)}}{r(r-1)}\dd r \equiv \int_{r_1}^{r_2}\frac{\sqrt{\Gamma(x)}}{x-1}\dd x
\end{equation}
which we use to fix $k$. We postpone the issue of writing down an explicit formula for $k$.
\medskip

We now solve~\ref{eq:eq_diff_wkb} with the new choice of $\Gamma(x)$ and we expand the solution in the near horizon and large distance regimes.

Equations~\ref{eq:uniform_solution_ansatz}, \ref{eq:def_uniform_V} and \ref{eq:uniform_smallness_condition} go through. We replicate the estimates for $\x$ at large $r$ and close to the horizon, obtaining
\begin{equation}
\left\{\begin{array}{llr}
\deriv{\x}{\r} \simeq \frac{\mu}{\sqrt{k}}, & \text{for } r\simeq1 \text{, thus} & \x\simeq\frac{\mu}{\sqrt{k}}\ln(r-1) \\
\sqrt{k}x\dd\x \simeq \mu r^{3/2} \dd\r, & \text{for } r\rightarrow\infty\text{, thus} & x\simeq 2\mu\sqrt{\frac{r}{k}}
\end{array}\right.
\end{equation}
where the algebraic identity $(1-r_1)(1-r_2)=1$ was used. Notice that, in neglecting subleading $\OO(1)$ terms, we are missing $\OO(1)$ corrections to the phase of the wave.

The solutions of~\ref{eq:eq_diff_wkb} are now
\begin{align}
e^{i \sqrt{k} \left(\log (x-1)-x\right)}\left\{c_1 U\left(a,b,2 i \sqrt{k} x-2 i \sqrt{k}\right)
+c_2 L_{n}^{c}\left(2 i \sqrt{k} x-2 i \sqrt{k}\right)\right\}
\end{align}
with
\begin{align}
a \eqdef \frac{1}{2} i \left(-i+2 \sqrt{k}-\sqrt{k} r_1+2 \sqrt{k} \sqrt{(r_1-1) (r_2-1)}-\sqrt{k} r_2\right)\\
b \eqdef 2 i \sqrt{k} \sqrt{(r_1-1) (r_2-1)}+1\\
n \eqdef -\frac{1}{2} i \left(-i+2 \sqrt{k}-\sqrt{k} r_1+2 \sqrt{k} \sqrt{(r_1-1) (r_2-1)}-\sqrt{k} r_2\right)\\
c \eqdef 2 i \sqrt{k} \sqrt{(r_1-1) (r_2-1)}
\end{align}
and $U(a,b,z)$ the confluent hypergeometric function and $L_n^a(z)$ the generalized Laguerre polynomial. The true solutions are obtained applying~\ref{eq:uniform_solution_ansatz} with $\x$ given above in the two relevant regimes.

\subsection{Near horizon}
\begin{equation}
\phi(r) = \frac{k^{1/4}}{\sqrt{\mu}} f(\x(r))
\end{equation}
Expanding $f$ for $r-1\ll1$, we obtain the sum of an infalling wave $(r-1)^{-i\mu}$ and an outgoing wave $(r-1)^{i\mu}$. Their coefficients must be set to 1 and 0 respectively.

Infalling:
\begin{equation}
\frac{c_1 2^{-2 i \sqrt{k}} e^{(\pi -i) \sqrt{k}} k^{\frac{1}{4}-i \sqrt{k}} \Gamma \left(2 i \sqrt{k}\right)}{\sqrt{\mu } \Gamma \left(\frac{1}{2} \left(1-i \sqrt{k} (\text{r1}+\text{r2}-4)\right)\right)}=1
\end{equation}
Outgoing:
\begin{equation}
\frac{k^{1/4}}{\sqrt{\mu }} e^{-i \sqrt{k}} \left(\frac{c_1\Gamma \left(-2 i \sqrt{k}\right)}{\Gamma \left(\frac{1}{2} \left(1-i \sqrt{k} (\text{r1}+\text{r2})\right)\right)}+c_2 L_{\frac{1}{2} \left(i \sqrt{k} (\text{r1}+\text{r2}-4)-1\right)}^{2 i \sqrt{k}}(0)\right)=0
\end{equation}

\subsection{Large $r$}
Similar expansions at $r\rightarrow\infty$ lead to the identification of an infalling wave $e^{-2i\mu\sqrt{r}}$ and an outgoing wave $e^{2i\mu\sqrt{r}}$. The leading term decays as $r^{-3/4}$ in both cases, as expected.

We implement the initial conditions just obtained and, after various intermediate steps, the end result simplifies to
\begin{align}
\label{eq:coefficienti_infinito_uniform_megacancro}
{\rm outgoing:} \: \frac{\sqrt{2} \sqrt[4]{k} \Gamma \left(-2 i \sqrt{k}\right) e^{\frac{\pi  \sqrt{k} \left(l^2-4 \mu ^2+l\right)}{4 \mu ^2}} \cosh \left(\frac{\pi  \sqrt{k} l (l+1)}{2 \mu ^2}\right) \Gamma \left(\frac{i l (l+1) \sqrt{k}}{2 \mu ^2}+\frac{1}{2}\right)}{\pi }\\
{\rm ingoing:}\;  \frac{\sqrt{2} \sqrt[4]{-k} e^{-\frac{\pi  \sqrt{k} l (l+1)}{4 \mu ^2}} \left(e^{\frac{\pi  \sqrt{k} l (l+1)}{\mu ^2}}+e^{4 \pi  \sqrt{k}}\right) \Gamma \left(\frac{1}{2}-\frac{i \sqrt{k} \left(l^2+l-4 \mu ^2\right)}{2 \mu ^2}\right)}{\left(e^{4 \pi  \sqrt{k}}-1\right) \Gamma \left(2 i \sqrt{k}+1\right)}
\end{align}
These expressions are the coefficients multiplying $\dfrac{1}{r^{3/4}}e^{\pm i \mu\sqrt{r}}$. The problem is thus completely solved if we can determine $k$, which is what we turn to now.

\subsection{$k$ for intermediate masses}
The integral of $\sqrt{V}$ defining~\ref{eq:def_uniform_k} is hard to do, but a robust and simple approximation can be achieved for $\mu\simeq\mu_c$, as defined in \ref{eq:critical_mass}. This amounts to $\mu\simeq\OO(1)$ for $l=1,2$, which are the most interesting cases.

To understand the simplification that occurs when $\mu\simeq\mu_c$, notice that the zeros $r_1,r_2$~\ref{eq:def_r12} of $V(r)$ collapse to $r_{1,2}=2$. Thus we can approximate the integral as follows
\begin{equation}
\int_{r_1}^{r_2} \frac{\mu\sqrt{(r-r_1)(r-r_2)}}{\sqrt{r}(r-1)}\dd r \simeq
\int_{r_1}^{r_2} \frac{\mu\sqrt{(r-r_1)(r-r_2)}}{\sqrt{2}}\dd r =
\frac{i \pi  \mu  (r_1-r_2)^2}{8 \sqrt{2}}
\end{equation}

The integral of $\sqrt{\Gamma}$ can be done, resulting in
\begin{equation}
\frac{\sqrt{k}}{2} i \pi  \left(r_1+r_2-4\right)
\end{equation}
Expanding both integrals for $\mu\simeq\mu_c$ one obtains
\begin{equation}
-2 i \sqrt{2} \pi  \left(\mu -\mu_c\right) = -\sqrt{k}\frac{8 i \pi  \left(\mu -\mu_c\right)}{\sqrt{l (l+1)}}\rightarrow
k = \frac{l(l+1)}{8}
\end{equation}
The weak link of this approximation is clearly the integral of $\sqrt{V}$. Computing it numerically, our approximation achieves 20\% accuracy even for masses as low as $\mu=0.3$ (for $l=1$), approximately 40\% of $\mu_c$. A simple scaling argument shows that the accuracy of our approximated integral is the same for all $l$ if $\mu/\mu_c(l)$ is given.%

\subsection{Conclusion}
For all $l$, plugging $k$ into the expressions~\ref{eq:coefficienti_infinito_uniform_megacancro} leads to great numerical accuracy to an incoming wave at infinity with amplitude $\sqrt{2}$, while the outgoing wave has amplitude $1$. We have no clear analytic understanding of this fact. As a consistency check, the current $J$ from section~\ref{sec:conserved_J} is conserved.
\smallskip

We developed an approximation for $\mu\simeq\mu_c$. It would be desirable to explore different mass ranges so as to cover the whole parameter space in figure~\ref{fig:validityofapprox}. However this is only meaningful if we are interested in very large angular momenta, which are phenomenologically less interesting.
\medskip

We now ask if this approximation is valid. While in section~\ref{subsec:large_mass} we could check the regime in which $\epsilon$ (equation~\ref{eq:uniform_smallness_condition}) is small, here we can only argue qualitatively.
\smallskip

The general lesson we learnt is that, besides matching the actual zeros of the differential equation, we should worry even when some terms simply {\it approach} zero. Looking at $V(r)$, this seems to happen only close to the horizon when $\mu\ll1$. This is the same regime which was causing trouble at $l=0$ in section~\ref{subsec:large_mass}.

If we trust that this regime is the only one where the uniform approximation breaks down, then we can simply argue that $\mu\ll1$ is a region of parameter space we already covered in section~\ref{subsec:small_mass}.

\section{Causality at the horizon}
\label{sec:appendix_wavepackets}
This appendix is dedicated to a more careful analysis of the causality condition when $\omega$ is not real.
We will assume stability, so that the imaginary part of $\omega$ must be negative $\rm{Im}(\omega)<0$ since $\phi\sim e^{-i\omega t}$ and we want a decaying solution.

Now as usual with complex frequencies, $\phi$ diverges near the horizon. Indeed, the approximate solution is
\begin{equation}
\label{eq:causal_component}
\phi\simeq (r-1)^{-i\omega}\sim (r-1)^{-i\Re(\omega)} (r-1)^{\rm{Im}(\omega)}
\end{equation}
while the non-causal solution is zero at the horizon
\begin{equation}
\label{eq:non-causal_component}
\phi\simeq (r-1)^{i\omega}\sim (r-1)^{i\Re(\omega)} (r-1)^{-\rm{Im}(\omega)}
\end{equation}
Hui and Baumann impose that the amplitude of the non-causal solution should be zero at the horizon. Since in this setup this is so automatically, does this mean that we can forget the causality condition? On the same note, should we be worried that the amplitude of the causal solution diverges at the horizon?
\medskip

Let us answer the question by schematically building a wave-packet out of the non-causal wave components~\ref{eq:non-causal_component} (the story for~\ref{eq:causal_component} is similar). We track the evolution of this (outgoing) wave packet in time and we ask what happens at very early times. If the wave packet vanishes close to the horizon causality is safe, otherwise it isn't.

We anticipate that the wave packet amplitude (at its peak) is almost unchanged for all $t$. This is so despite its wave components~\ref{eq:non-causal_component} approaching zero for $r\rightarrow r_{\rm{BH}}$ at any given time $t$.
Thus we want to set the non-causal solution to zero, otherwise an observer hovering outside the horizon would see stuff flying out. A free falling observer would instead perceive this as an infinite energy solution.
\bigskip

Now we carry out the analysis sketched above. The wavepacket is a superposition of~\ref{eq:non-causal_component} with modulation $\tilde{\phi}(\omega)$. We are neglecting the infrared (large $r$) deformations of~\ref{eq:non-causal_component}, but they are irrelevant for the present purpose.
\begin{equation}
\phi(r,t) = \int\dd\omega\, e^{i\omega\ln(r-r_{\rm{BH}})} e^{-i\omega t} \tilde{\phi}(\omega)
\end{equation}
As an example, we choose $\tilde{\phi}(\omega) = e^{-i\omega \ln r_0}f_{\omega_0}(\omega)$. The exponential prefactor translates the solution at time $t=0$. $f$ is some real function peaked at $\omega_0$, which gives the size of the wave packet.

We split real and imaginary part of $\omega$ into $\omega\rightarrow\omega-i\epsilon$
\begin{equation}
\phi(r,t) = e^{\epsilon\ln(r-r_{\rm{BH}})-\epsilon t-\epsilon\ln r_0} \int\dd\omega\, e^{i\omega\ln(r-r_{\rm{BH}})} e^{-i\omega t} e^{-i\omega \ln r_0}f_{\omega_0}(\omega)
\end{equation}
and we ask when is the integral large. A saddle point approximation yields the position of the peak of the wavefunction in time:
\begin{align}
\partial_\omega\left(\omega(\ln(r-r_{\rm{BH}})-t-\ln r_0)\right)=0\rightarrow
\ln(r-r_{\rm{BH}})-t-\ln r_0 = 0\\
r(t) = r_{\rm{BH}} + r_0 e^{t}
\end{align}
The wavepacket flies out of the horizon as predicted, and its initial position is controlled by $r_0$ as initially claimed. Let's track the wave packet (peak) amplitude in time. It is defined as
\begin{equation}
A(t) \eqdef \phi(r(t),t) \sim e^{\epsilon\ln(r(t)-r_{\rm{BH}})-\epsilon t-\epsilon\ln r_0} =  1
\end{equation}
so it remains finite (even constant) at all times. While constancy is an artefact of the near horizon limit (i.e. the amplitude does change at large $t$), finiteness is robust and implies that the wavepacket is non-zero arbitrarily close to the horizon.

Similarly, a wave packet build with the causal waves~\ref{eq:causal_component} will also remain finite as it approaches the horizon, despite the individual waves blowing up at $r=r_{\rm{BH}}$.

\section{Soliton bound states: energy levels}
\label{app:soliton_frequency}
In this appendix we try to analytically compute the spectrum of the soliton. We will be able to compute the factor appearing in $\mu^2 GM_{\rm{sol}} r_{\rm{sol}}$ and upper bound $C^2$. The precise value of $C^2$ was however too difficult for us to compute in a fully analytic approach.
We discuss the case with no black hole inside the soliton, but the more complicated case \text{soliton+black hole} is expected to be qualitatively similar.
\bigskip

The first point we want to make is that $C$ is much smaller than unity in physically realistic situations, again because of the stability of the soliton.
\smallskip

While $C^2=1-(1+\gamma)^2$ dropped out of all equations following~\ref{eq:chi_large_distance}, 
the explicit form of $\Phi$ in equation~\ref{eq:Phi} enables us to compute $C^2$. Recall that $r_{\rm{sol}}$ is defined as the (single) zero of $-2\Phi(r)-C^2=0$, so
\begin{equation}
-2\frac{-GM_{\rm{sol}}}{r_{\rm{sol}}}-C^2=0\rightarrow C^2 = \frac{2GM_{\rm{sol}}}{r_{\rm{sol}}}
\end{equation}
The newtonian limit of equation~\ref{eq:eq_diff_selfG} has a well known scale invariance that sends a solution $\left(r_{\rm{sol}},M_{\rm{sol}},C^2\right)$ into one with $\left(\dfrac{r_{\rm{sol}}}{\lambda},\lambda M_{\rm{sol}},\lambda^2C^2\right)$, so the new solution is more squeezed and massive. For $\lambda$ large enough we expect two a priori unrelated things: first the newtonian gravity approximation should break down and, possibly at even larger $\lambda$, the soliton itself might become unstable. Stability forces us to require
\begin{equation}
2GM_{\rm{sol}}\ll r_{\rm{sol}}\rightarrow C\ll 1
\end{equation}
Many authors~\cite{Schive:2014dra,Schive:2014hza,bar_galactic_2018,Marsh:2015wka} set $\chi|_{r=0}=1,\mu=1$ and numerically find $\gamma\simeq0.69$ ($\rightarrow C\simeq0.9$), which thus must be interpreted not as a physically meaningful solution, but rather as a seed we can use to produce solutions with much smaller $C$. $\gamma\simeq1$ further implies $\lambda\ll1$, suggesting that the newtonian limit should always be valid (at least outside $r_{\rm{e}}$).
\medskip

After these preliminaries, we now discuss the quantization condition. We already mentioned that the requirement is
\begin{equation}
\cos\left(\mu\int^r_{r_{\rm{sol}}}\sqrt{-2\Phi-C^2}\dd r+\frac{\pi}{4}\right)\equiv \sin\left(\mu\int^r_{0}\sqrt{-2\Phi-C^2}\dd r\right)
\end{equation}
in order to have $\chi$ bounded at the origin and matched to an asymptotically decaying solution at infinity.
This means that
\begin{equation}
\mu\int^{r_{\rm{sol}}}_0\sqrt{-2\Phi-C^2}\dd r = \frac{\pi}{2}\left(\frac{3}{2}+2n\right)
\end{equation}
Substituting the value of $\Phi$ obtained in~\ref{eq:Phi} we can do the integral and obtain
\begin{equation}
\mu\sqrt{GM_{\rm{sol}}r_{\rm{sol}}} (0.95) = \frac{\pi}{2} \left(\frac{3}{2}+2n\right)
\end{equation}
The first conclusion is that the approximate relation $\mu^2GM_{\rm{sol}}r_{\rm{sol}}\simeq1$ found in~\ref{eq:soliton_size} using a scaling argument is essentially correct, but it can now be made more precise. Reinserting $\hbar,c$ we obtain for the lowest energy level $n=0$
\begin{equation}
\label{eq:GMsrsmu2_precise}
GM_{\rm{sol}}r_{\rm{sol}}\mu^2\simeq (6.2),\qquad
M_{\rm{sol}}r_{\rm{sol}}\simeq (5.3) \cdot 10^5\left(\frac{\mu^2}{10^{-19}\si{eV}}\right)^{-2} M_{\astrosun}\si{pc}
\end{equation}
which is roughly twice the value reported in~\cite{bar_looking_2019,bar_galactic_2018}. The discrepancy could be due to the approximations we have made. Moreover, we have a prediction for the mass scale of all excited bound states $n>0$
\begin{equation}
M_{\rm{sol}}r_{\rm{sol}}\simeq (2.4)\cdot10^5\left(\frac{\mu^2}{10^{-19}\si{eV}}\right)^{-2}\left(\frac{3}{2}+2n\right)^2 M_{\astrosun}\si{pc}
\end{equation}
\medskip

Additionally, Bar~\cite{bar_galactic_2018} has a prediction for what $\gamma,M_{\rm{sol}},r_{\rm{sol}}$ should be if we set $\chi=1$ at the origin (so that the scaling symmetry is broken). We will now try to reproduce their numerical results.

First, $M_{\rm{sol}}$ and $r_{\rm{sol}}$ are related by the already mentioned relation $GM_{\rm{sol}}r_{\rm{sol}}\mu^2\simeq1$, so we will only determine $r_{\rm{sol}}$. We start from $\chi(0)\equiv\chi_0=1$, which implies
\begin{equation}
    4\pi\rho(0) = \frac{4\pi \mu^2 \chi_0^2}{4\pi G} \simeq (8.3) \frac{M_{\rm{sol}}}{r_{\rm{sol}}^3}
\end{equation}
where the last step comes from equation~\ref{eq:rho_full}. This implies $GM_{\rm{sol}}\simeq\frac{\mu^2 r_{\rm{sol}}^3}{(8.3)}$ which we can use together with equation~\ref{eq:GMsrsmu2_precise} to deduce
\begin{equation}
    (\mu r_{\rm{sol}})^4\simeq (6.2) (8.3)\rightarrow \mu r_{\rm{sol}}\simeq 2.7
\end{equation}
Bar has $\mu r_{\rm{sol}}\simeq 1.3$ in~\cite{bar_galactic_2018}, so again we are qualitatively correct. It also makes sense that for ground state of the soliton to have size roughly a few compton wavelengths of the scalar particle. The scale parameter $\lambda$, which we can define as $\chi_0=\lambda^2$, thus has the further interpretation that $\lambda^{-1}$ counts the number of oscillations of the field inside the soliton.

Finally, we now turn to the computation of $C^2$. From $C^2=\frac{2GM_{\rm{sol}}}{r_{\rm{sol}}}$, we can deduce
\begin{equation}
    C_{us}^2\simeq 2\frac{(\mu r_{\rm{sol}})^2}{(8.3)}\simeq 1.7,\quad C_{Bar}^2\simeq0.4
\end{equation}
where we first used our own $\mu r_{\rm{sol}}\simeq 2.7$ and then we used Bar's $\mu r_{\rm{sol}}\simeq1.3$. To compare with the value reported in~\cite{Chavanis:2011zi,Chavanis:2011zm,Marsh:2015wka}, we should make the (improper) identification $C^2\equiv-2\gamma$, which brings our result to $-\gamma\simeq 1.7/2=0.85$, not far from $0.69$ obtained numerically.

\printbibliography[heading=bibintoc]

\end{document}